\documentstyle[aps,psfig]{revtex}
\newcommand{\T}{{\large$t$}}
\newcommand{\fw}{\!\!/}
\newcommand{\fd}{\!\!\!/}

\begin{document}
\draft
\preprint{MKPH-T-97-27}
\title{Multipole analysis of pion photoproduction based on fixed \T 
       \hspace{3mm}dispersion relations and unitarity}

\author{O.~Hanstein, D.~Drechsel and L.~Tiator}
\address{Institut f\"ur Kernphysik, Universit\"at Mainz, 55099 Mainz, Germany}
\date{\today}
\maketitle

\begin{abstract}
  We have analysed pion photoproduction imposing constraints from
  fixed $t$ dispersion relations and unitarity. Coupled integral
  equations for the $S$ and $P$ wave multipoles were derived from the
  dispersion relations and solved by the method of Omn\`es and
  Muskhelishvili. The free parameters were determined by a fit to the
  most recent data for $\pi^{+}$ and $\pi^{0}$ production on the
  proton as well as $\pi^{-}$ production on the neutron, in the energy
  range 160 MeV $\leq E_{\gamma} \leq$ 420 MeV. The lack of high
  precision data on the neutron and of polarization observables leads
  to some limitations of our results. Especially the multipole
  $M_{1-}$ connected with the Roper resonance $P_{11}$(1440) cannot be
  determined to the required precision. Our predictions for the
  threshold amplitudes are in good agreement with both the data and
  chiral perturbation theory. In the region of the $\Delta(1232)$ we
  have determined the ratio of electric quadrupole and magnetic dipole
  excitation. The position of the resonance pole is obtained in
  excellent agreement with pion-nucleon scattering, and the complex
  residues of the multipoles are determined with the speed-plot
  technique.  

\pacs{PACS numbers: 13.60.Le, 14.20.Gk, 11.55.Fv, 11.80.Et \\ 
{\em Keywords}: Pion photoproduction, dispersion relations, 
 partial-wave analysis }

\end{abstract}

\section{Introduction}

Pion photoproduction was extensively studied both theoretically and 
experimentally in the sixties and seventies. The basic theoretical approach 
was given by dispersion theory, especially by fixed $t$ dispersion relations. 
Since perturbative methods of quantum field theory fail in the case of 
the strong interactions in the resonance region, dispersion 
relations seemed to be the only possible 
tool to treat hadronic processes on a quantitative level. 
Given the technical possibilities of that time, the statistics of the data 
was moderate and systematical errors were large. With the development
of quantum chromodynamics (QCD) as a fundamental field theory 
in terms of quarks and gluons, the theoretical understanding of strong 
interactions has improved at the higher energies. However our understanding 
of hadron physics in the nonperturbative region at energies of the order of 
1 GeV remains still unsatisfactory. 
Although quark models of different kinds give some description 
of the baryon and meson spectra, their predictive power 
for reactions is still quite poor. The only approach to infer all the 
symmetries of QCD into the physics at these energies 
is given by chiral perturbation theory (ChPT). However, the 
results of ChPT can be applied only in the threshold region. Recently, it has 
been investigated how to include the most prominent resonance of the 
nucleon, the $\Delta$(1232), within the framework of heavy baryon 
ChPT. However, the first results achieved by this technique are not 
yet satisfactory. 
On the other side modern electron accelerators have provided a 
host of high precision data in the 90's, and will continue to do so. Beams 
of high current and high duty factor together with considerably improved 
particle detection techniques have reduced the statistical errors to 
the order of a few percent, and promise to keep control of the systematical 
errors at the same level. To interpret these data with respect to the most 
interesting features, i.e. the threshold behaviour and the electromagnetic 
excitation of resonances, a partial wave analysis is mandatory. To ensure 
the consistency and uniqueness of such an analysis, constraints from unitarity 
and dispersion relations have to be imposed. Such concepts have proven 
to be quite successful 
in pion-nucleon scattering \cite{Hoe83}. In comparison with that field, 
the situation in pion photoproduction is considerably more complex. The 
spin and isospin structure leads to twelve independent amplitudes, while 
in pion-nucleon scattering there are only four such amplitudes. As a  
consequence a complete experiment \cite{Bar75} requires the use 
of many polarization observables. Such a complete  experiment has not 
yet been performed. However, the 
new experiments provide an ever increasing amount of precise and new data. 
At present, the experimental thrust is mainly on measurements 
near threshold and around the $\Delta$(1232) resonance. In the coming 
years, a series of experiments at Jefferson Lab will cover the whole resonance 
region. Restricting our theoretical investigations to the threshold region 
and the low-lying resonances, we are lead to choose 
the method of Omn\`es and Muskhelishvili to 
analyse the existing data, because it introduces a natural parametrization and 
fulfills the constraints of unitarity at the same trace. 

In the future, the required shift to include also higher energies will make 
it necessary to take account of the full content of dispersion methods and 
to apply more involved techniques. At the moment, however, we obtain an 
excellent representation of the analysed data by use of the method of 
Ref.~\cite{Sch69}. Therefore we deem it worthwhile extending this 
method to the second resonance region as a next step.

In Sec.~2 of this contribution we briefly recall the kinematics, amplitudes 
and (polarization) observables of pion photoproduction. The essential 
ingredients of dispersion relations at fixed $t$ are outlined in Sec.~3. The 
method of Omn\`es and Muskhelishvili to solve these coupled integral equations 
is given in Sec.~4. In particular, we also discuss our choice of the free 
parameters of that method, which are then fitted to our data basis 
(Sec.~5). This fit defines the result for the multipole amplitudes, which 
are presented in Sec.~6. In particular we give our prediction for the 
threshold production of charged and neutral pions, the ratio of 
electric quadrupole and magnetic dipole radiation in the region of the 
$\Delta$(1232), and the inclusive cross sections. We conclude with a 
short summary and outlook in Sec.~7.

\section{Formalism}

\subsection{Kinematics and conventions}

The four-momenta occuring in the reaction are denoted by
$q = (\omega,\vec q)$ for the photon,  
$k = (\omega_{\pi},\vec k)$ for the pion, $P_{i} = (P_{i}^{0},\vec P_{i})$ 
and $P_{f} = (P_{f}^{0},\vec P_{f})$ for the nucleon in the initial 
and the final state, respectively. With this notation 
the conventional Mandelstam variables are
\begin{eqnarray}
  \label{mandel-var}
  s & = & (P_{i} + q)^{2} = (P_{f} + k)^{2} = W^{2}, \\
  t & = & (k - q)^{2} = (P_{f} - P_{i})^{2}, \\
  u & = & (P_{f} - q)^{2} = (P_{i} - k)^{2},
\end{eqnarray}
where $W$ is the {\it cm} energy. Since all 
calculations are performed in the {\it cm} frame, we further introduce 
the {\it cm} energy of the nucleons,
\begin{equation}
E_{i,f} = P_{i,f}^{0,cm}.
\end{equation}
In the following we will suppress the index {\it cm} for the momenta,
\begin{eqnarray}
\vec q & = & \vec q^{\hspace{1mm}cm} = -\vec P_{i}^{cm}, \\
\vec k & = & \vec k^{cm} = -\vec P_{f}^{cm},
\end{eqnarray} 
and introduce the quantity $x = \cos\theta$ 
instead of the scattering angle $\theta = \theta^{cm}$. 
In the discussion of our results for the amplitudes and observables, the photon 
energy is usually given in the {\it lab} frame, 
$E_{\gamma} = E_{\gamma}^{lab}$.

\subsection{Amplitudes}

The invariant amplitudes $A_{i}(s,t,u)$ are defined according to 
\cite{Che57}. The relativistic invariants are\footnote{For the metric and 
the $\gamma$ matrices we adopt the conventions of Ref.~\protect\cite{Bjo64}.}
\begin{eqnarray}
  \label{M_i}
  M_{1} & = & i\gamma_{5}\hspace{0.5mm}\varepsilon\fw \hspace{0.5mm}q\fd,\\
  M_{2} & = & 2i\gamma_{5}(P\cdot q \hspace{1mm}k\cdot\varepsilon 
               - P\cdot\varepsilon \hspace{1mm}k\cdot q),\\
  M_{3} & = & i\gamma_{5}(\varepsilon\fw \hspace{1mm}k\cdot q 
             - q\fd \hspace{1mm}k\cdot\varepsilon),\\
  M_{4} & = & 2i\gamma_{5}(\varepsilon\fw\hspace{0.5mm} P\cdot q 
                - q\fd \hspace{0.5mm}P\cdot\varepsilon 
               - m_{N}\hspace{0.5mm}\varepsilon\fw \hspace{0.5mm}q\fd),
\end{eqnarray}
where $\varepsilon$ is the polarization vector of the photon and 
$P = \frac{1}{2}(P_{i} + P_{f})$. The scattering operator $T$, to be
evaluated between the Dirac spinors of the nucleon, is expressed in terms of 
the invariants by
\begin{equation}
  \label{T_inv}
  T = \sum_{i=1}^{4}A_{i}(s, t, u)M_{i}.
\end{equation}
The photoproduction amplitudes $A_{i}$ can be decomposed in isospace 
according to
\begin{equation}
  \label{iso_komp}
  A_{i} = \frac{1}{2}A_{i}^{(-)}[\tau_{\alpha},\tau_{0}] 
          + A_{i}^{(+)}\delta_{\alpha 0} + A_{i}^{(0)}\tau_{\alpha},
\end{equation}
with the Pauli matrices $\tau_{\alpha}$ acting on the isospinor of the 
nucleon. The behaviour of the invariant amplitudes under crossing is
\begin{equation}
  \label{crossing}
  A_{k}^{I}(s,t,u) = \varepsilon^{I}\xi_{k}A_{k}^{I}(u,t,s),
\end{equation}
\begin{displaymath}
  I=+,0,-;\hspace{0.5cm}\xi_{1}=\xi_{2}=-\xi_{3}=\xi_{4}=1;
  \hspace{0.5cm}\varepsilon^{+}=\varepsilon^{0}=-\varepsilon^{-}=1.
\end{displaymath}

The physical amplitudes may be expressed by the isospin combinations
\begin{eqnarray}
  \label{iso_phys_1}
  A(\gamma p \rightarrow n\pi^{+}) & = & \sqrt{2}(A^{(-)}+A^{(0)})
  = \sqrt{2}(_{p}A^{(\frac{1}{2})}-\frac{1}{3}A^{(\frac{3}{2})}), \\
  A(\gamma p \rightarrow p\pi^{0}) & = & A^{(+)}+A^{(0)}
  = _{p}A^{(\frac{1}{2})}+\frac{2}{3}A^{(\frac{3}{2})}, \\
  A(\gamma n \rightarrow p\pi^{-}) & = & -\sqrt{2}(A^{(-)}-A^{(0)})
  = \sqrt{2}(_{n}A^{(\frac{1}{2})}+\frac{1}{3}A^{(\frac{3}{2})}), \\
  \label{iso_phys_4}
  A(\gamma n \rightarrow n\pi^{0}) & = & A^{(+)}-A^{(0)}
  = -_{n}A^{(\frac{1}{2})}+\frac{2}{3}A^{(\frac{3}{2})}.
\end{eqnarray}
The amplitudes $A^{(\frac{1}{2})}$ and $A^{(\frac{3}{2})}$ refer to final 
states of definite isospin ($\frac{1}{2}$ or $\frac{3}{2}$). They are 
combinations of the quantities $A^{(+)}$ and $A^{(-)}$ of Eq.~(\ref{iso_komp}),
\begin{equation}
  \label{iso_iso_13}
  A^{(\frac{1}{2})} = A^{(+)} + 2 A^{(-)},\hspace{3mm}
  A^{(\frac{3}{2})} = A^{(+)} - A^{(-)}.
\end{equation}
For the isospin $\frac{1}{2}$ channel it is also useful to define the 
combinations
\begin{equation}
  \label{iso_iso_pn}
  _{p}A^{(\frac{1}{2})} = A^{(0)} + \frac{1}{3}A^{(\frac{1}{2})},
\hspace{3mm}
  _{n}A^{(\frac{1}{2})} = A^{(0)} - \frac{1}{3}A^{(\frac{1}{2})},
\end{equation}
which enter into the amplitudes of the reactions on the proton or the 
neutron, respectively.

Evaluating Eq.~(\ref{T_inv}) between the nucleon 
Dirac spinors in the {\it cm} frame, expressing the result 
by Pauli spinors and multiplying by a factor 
$m_{N}/4\pi W$, we obtain the form \cite{Che57}
\begin{equation}
  \label{CGLN}
  {\cal F} = i\vec\sigma\cdot\vec\varepsilon\hspace{0.5mm}{\cal F}_{1} 
                +\vec\sigma\cdot\hat k\hspace{0.5mm}\vec\sigma\cdot
                (\hat q\times\vec\varepsilon){\cal F}_{2}
                +i\vec\sigma\cdot\hat q\hspace{0.5mm}\hat k\cdot
                \vec\varepsilon{\cal F}_{3}
                +i\vec\sigma\cdot\hat k\hspace{0.5mm}\hat k\cdot
                \vec\varepsilon{\cal F}_{4},
\end{equation}
where $\hat q$ and $\hat k$ are unit vectors in the direction of the 
respective momenta. The {\it cm} differential cross section can be expressed 
through the CGLN amplitudes ${\cal F}_{i}$ defined by Eq.~(\ref{CGLN}),
\begin{equation}
  \label{dsig}
  \frac{d\sigma}{d\Omega} = \frac{\mid\vec k\mid}{\mid\vec q\mid}
  \mid\chi_{f}^{\dag}{\cal F}\chi_{i}\mid^{2},
\end{equation}
where $\chi_{f,i}$ are nucleon spinors. 
The connection between the CGLN amplitudes and the invariant amplitudes is
given by
\begin{eqnarray}
  \label{A_F_Gl}
  F_{1} & = & 4\pi\frac{2W}{(W-m_{N})}\hspace{0.5mm}
      \frac{{\cal F}_{1}}{[(E_{f}+m_{N})(E_{i}+m_{N})]^{\frac{1}{2}}} 
      \nonumber \\
        & = & A_{1} + (W-m_{N})A_{4}
           -\frac{t-m_{\pi}^{2}}{2(W-m_{N})}(A_{3}-A_{4}), \nonumber \\
  F_{2} & = & 4\pi\frac{2W}{(W-m_{N})}\hspace{0.5mm}
            \left(\frac{E_{f}+m_{N}}{E_{i}+m_{N}}\right)^{\frac{1}{2}} 
       \frac{{\cal F}_{2}}{\mid\vec k\mid} \nonumber \\
        & = &  -A_{1} + (W+m_{N})A_{4} 
           - \frac{t-m_{\pi}^{2}}{2(W+m_{N})}(A_{3}-A_{4}), \\
  F_{3} & = & 4\pi\frac{2W}{(W-m_{N})} \hspace{0.5mm}
      \frac{{\cal F}_{3}}{[(E_{f}+m_{N})(E_{i}+m_{N})]^{\frac{1}{2}}
        \mid\vec k\mid}  \nonumber \\
        & = & (W-m_N)A_{2} + (A_{3}-A_{4}), \nonumber \\
  F_{4} & = & 4\pi\frac{2W}{(W-m_{N})}\hspace{0.5mm}
            \left(\frac{E_{f}+m_{N}}{E_{i}+m_{N}}\right)^{\frac{1}{2}} 
       \frac{{\cal F}_{4}}{\mid\vec k\mid^{2}} \nonumber \\
        & = & -(W+m_N)A_{2} + (A_{3}-A_{4}). \nonumber
\end{eqnarray}
For practical calculations it is useful to introduce a matrix form for 
the different amplitudes,
\begin{equation}
  \label{Mat_F_A}
  \vec{\cal {F}} = {\mbox{\bf H}}\vec F,\hspace{3mm}
  \vec F = {\mbox{\bf C}}\vec A, \hspace{3mm}
  \vec A = {\mbox{\bf C}}^{-1}\vec F,
\end{equation}
where $\vec A$, $\vec{\cal {F}}$ and $\vec F$ are vectors with the
four elements defined in Eqs.~(\ref{T_inv}), (\ref{CGLN}) and
(\ref{A_F_Gl}) respectively.  The $4\times 4$ matrices $\mbox{\bf{C}}$
  and $\mbox{\bf{H}}$ contain kinematical factors only and are given
  in appendix \ref{matrices}.

The CGLN amplitudes are expanded into 
the multipoles $E_{l\pm}$ and $M_{l\pm}$ which conserve parity 
and total angular momentum $J$,
\begin{eqnarray}
  \label{CGLN_mul}
  {\cal F}_{1}(W, \theta) & = & \sum_{l \geq 0}\{(lM_{l+}(W)
                       +E_{l+}(W))P'_{l+1}(x) \nonumber \\
  & & \hspace{0.8cm} + [(l+1)M_{l-}(W)+E_{l-}(W)]P'_{l-1}(x)\}, 
           \nonumber \\[0.4cm]
  {\cal F}_{2}(W, \theta) & = & \sum_{l\geq 1}[(l+1)M_{l+}(W)
                                  +lM_{l-}(W)]P'_{l}(x), \nonumber \\
  {\cal F}_{3}(W, \theta) & = & \sum_{l\geq 1}
                                 [(E_{l+}(W)-M_{l+}(W))P''_{l+1}(x) \\
  & & \hspace{0.8cm}           +(E_{l-}(W)+M_{l-}(W))P''_{l-1}(x)], 
           \nonumber \\[0.4cm]
  {\cal F}_{4}(W, \theta) & = & \sum_{l\geq 2}
                            (M_{l+}(W)-E_{l+}(W)-M_{l-}(W)-E_{l-}(W))P''_{l}(x).
  \nonumber
\end{eqnarray}
$E$ and $M$ denote the electric or magnetic character of the incoming photon, 
the indices $l\pm$ describe the coupling of the pion angular momentum $l$ 
and the nucleon spin to the total angular momentum  $J = l \pm\frac{1}{2}$. 

The 4 partial waves with angular momentum $l$ may be written in terms of 
a vector, 
\begin{equation}
  \label{Vec_Mul}
  \vec{\cal {M}}_{l}(W)^{\mbox{T}} = \left(
    E_{l+}(W), E_{l-}(W), M_{l+}(W), M_{l-}(W) \right), 
\end{equation}
where T stands for transposed. 
The matrix form of the multipole expansion Eq.~(\ref{CGLN_mul}) 
and its inverse read
\begin{eqnarray}
  \label{Mat_mul}
  \vec{\cal {F}}(W, x) & = & 
   \sum_{l = 0}^{\infty}{\mbox{\bf G}}_{l}(x)\vec{\cal {M}}_{l}(W), \\
  \label{Mat_proj}
  \vec{\cal {M}}_{l}(W) & = & \int\limits_{-1}^{1}
                        dx{\mbox{\bf D}}_{l}(x)\vec{\cal F}(W, x),
\end{eqnarray}
with the $4\times 4$ matrices ${\mbox{\bf G}}_{l}$ and ${\mbox{\bf D}}_{l}$ given in 
appendix \ref{matrices}.

The multipole amplitudes ${\cal M}_{l\pm}$ are complex functions 
of the {\it cm} energy $W$. Below inelastic threshold, the 
Fermi-Watson theorem \cite{Wat54} allows one to express the complex phases 
of the multipoles by the 
corresponding pion-nucleon scattering phase shifts,
\begin{equation}
  \label{multi_wat}
  {\cal M}_{l\pm}^{I}(W) 
   = \mid{\cal M}_{l\pm}^{I}(W)\mid e^{i(\delta_{l\pm}^{I}(W)+n\pi)},
\end{equation}
with $I = \frac{1}{2}, \frac{3}{2}$.

\subsection{Observables}
If polarization degrees of freedom are taken into account, 
the differential cross section of pion photoproduction 
takes the form \cite{Dre92}
\begin{eqnarray}
  \label{d_sig}
  \frac{d\sigma}{d\Omega} & = & \frac{\mid\vec k\mid}{\mid\vec q\mid}\{
  (R_{T}+P_{n_{i,f}}R_{T}^{n_{i,f}})+\Pi_{T}[(R_{TT}
  +P_{n_{i,f}}R_{TT}^{n_{i,f}})\cos 2\varphi 
  \\
  & & -(P_{l_{i,f}}R_{TT}^{l_{i,f}}+P_{t_{i,f}}R_{TT}^{t_{i,f}})\sin 2\varphi] 
  + \Pi_{C}(P_{l_{i,f}}R_{TT'}^{l_{i,f}} + P_{t_{i,f}}R_{TT'}^{t_{i,f}})\}.
\nonumber
\end{eqnarray}
The various response functions $R$ may be expressed 
through the CGLN amplitudes. 
$P_{i}$ are the components of the nucleon polarization vector, 
$\Pi_{T}$ is the degree of linear polarization of the photon, $\varphi$ 
the angle between the reaction plane and the photon polarization, and 
$\Pi_{C}$ the degree of circular photon polarization. The observables
containing the information of the various polarization 
measurements are the appropriate 
asymmetries. In the following we will restrict the discussion 
to single polarization observables only.  
The beam asymmetry $\Sigma$ which is measured with linearly 
polarized photons, perpendicular and parallel to the production plane is
\begin{equation}
  \label{Sigma}
  \Sigma = \frac{d\sigma^{\perp}-d\sigma^{\parallel}}
                {d\sigma^{\perp}+d\sigma^{\parallel}}
         =\frac{-R_{TT}}{R_T}.
\end{equation}
The target asymmetry $T$ and the recoil polarization $P$, 
\begin{equation}
  \label{Targetasym}
  T = \frac{d\sigma^{P_{n_{i}}=1}-d\sigma^{P_{n_{i}}=-1}}
                {d\sigma^{P_{n_{i}}=1}+d\sigma^{P_{n_{i}}=-1}}
         =\frac{R_{T}^{n_{i}}}{R_T},
\end{equation}
\begin{equation}
  \label{recoilpol}
  P = \frac{d\sigma^{P_{n_{f}}=1}-d\sigma^{P_{n_{f}}=-1}}
                {d\sigma^{P_{n_{f}}=1}+d\sigma^{P_{n_{f}}=-1}}
         =\frac{R_{T}^{n_{f}}}{R_T}, 
\end{equation}
are determined with respect to the polarization of the incoming (outgoing) 
nucleon perpendicular to the production plane. In appendix \ref{resp_func} 
the relevant response functions are given as 
functions of the CGLN amplitudes.

\section{Dispersion relations at fixed \T}

The starting point of our analysis are dispersion relations 
at fixed $t$ for the invariant amplitudes \cite{Bal61},
\begin{equation}
  \label{fixed_t_photo}
  {\mbox{Re}}A_{k}^{I}(s, t) = A_{k}^{I,\mbox{pole}}(s, t)+
  \frac{1}{\pi}{\cal P}\!\!\int\limits_{s_{\mbox{thr}}}^{\infty}
  ds'\left(\frac{1}{s'-s}+\frac{\varepsilon^{I}\xi_{k}}{s'-u}\right)
  {\mbox{Im}}A_{k}^{I}(s', t),
\end{equation}
with $I=0,\pm$ and $s_{\mbox{thr}} = (m_{N}+m_{\pi})^{2}$.
The pole term contributions $A_{k}^{I,\mbox{pole}}$ can be obtained 
by evaluating the Born approximation in pseudoscalar coupling,
\begin{eqnarray}
  \label{inv_pol}
  A_{1}^{I,\mbox{pole}}(s,t) & = & \frac{eg}{2}\left(\frac{1}{s-m_{N}^{2}}
        +\frac{\varepsilon^{I}}{u-m_{N}^{2}}\right), \nonumber \\
  A_{2}^{I,\mbox{pole}}(s,t) & = & -\frac{eg}{t-m_{\pi}^{2}}
           \left(\frac{1}{s-m_{N}^{2}}
          +\frac{\varepsilon^{I}}{u-m_{N}^{2}}\right)
       =  \frac{eg}{s-m_{N}^{2}}\left(\frac{\varepsilon^{I}-1}{t-m_{\pi}^{2}}
           +\frac{\varepsilon^{I}}{u-m_{N}^{2}}\right),\\
  A_{3}^{I,\mbox{pole}}(s,t) & = & 
         -\frac{eg}{2m_{N}}\frac{\kappa^{I}}{2}\left(
          \frac{1}{s-m_{N}^{2}}-\frac{\varepsilon^{I}}{u-m_{N}^{2}}\right),
  \nonumber \\
  A_{4}^{I,\mbox{pole}}(s,t) & = & 
         -\frac{eg}{2m_{N}}\frac{\kappa^{I}}{2}\left(
          \frac{1}{s-m_{N}^{2}}+\frac{\varepsilon^{I}}{u-m_{N}^{2}}\right),
  \nonumber
\end{eqnarray}
  \label{g_I_def}
with $\kappa^{(+,-)}=\kappa_{p}-\kappa_{n}$ and 
$\kappa^{(0)}=\kappa_{p}+\kappa_{n}$,
where $\kappa_{p}$ and $\kappa_{n}$ are the anomalous magnetic moments 
of the proton and the neutron, respectively.
For the pion nucleon coupling constant $g$ we take the value 
$g^{2} / 4\pi = 14.28$ \cite{Hoe83}.

The next step is to apply the multipole projection to the dispersion 
relations (\ref{fixed_t_photo}). With the help of Eqs. (\ref{Mat_F_A}) 
we obtain
\begin{equation}
  \label{disp_F}
  {\mbox Re}{\mbox{\bf F}}^{I}(W,x) = {\mbox{\bf F}}^{I,\mbox{pole}}(W,x)
  +\frac{1}{\pi}
  {\cal P}\!\!\int\limits_{W_{\mbox{thr}}}^{\infty}dW'
  {\mbox{\bf E}}^{I}(W, W', x'){\mbox Im}{\mbox{\bf F}}^{I}(W', x'),
\end{equation}
where $x' = \cos\theta'$ corresponds to $W'$ at fixed $t$.
With
\begin{equation}
  \label{t_mpi}
  t - m_{\pi}^{2}=-2\omega\omega_{\pi}+2\mid\vec k\mid\mid\vec q\mid x
                 =-2\omega'\omega_{\pi}'+2\mid\vec k'\mid
                   \mid\vec q\hspace{1mm}'\mid x'
\end{equation}
we have
\begin{equation}
  \label{x_s}
  x'= ex + f;\hspace{0.5cm}e=\frac{\mid\vec k\mid\mid\vec q\mid}
                                  {\mid\vec k'\mid\mid\vec q\hspace{1mm}'\mid};
\hspace{0.5cm}f = \frac{-\omega\omega_{\pi}+\omega'\omega_{\pi}'}
{\mid\vec k'\mid\mid\vec q\hspace{1mm}'\mid}.
\end{equation}
The entries of the matrix ${\mbox{\bf E}}^{I}(W, W', x')$ are given in  appendix 
\ref{matrices}.
Application of  Eqs.~(\ref{Mat_mul}) and (\ref{Mat_proj}) leads to a  
system of coupled integral equations for the multipole amplitudes,
\begin{equation}
  \label{multi_int}
  {\mbox{Re}}{\cal M}_{l}^{I}(W) =
   {\cal M}_{l}^{I,\mbox{pole}}(W) + \frac{1}{\pi}{\cal P}
  \!\!\int\limits_{W_{\mbox{thr}}}^{\infty}dW'\sum_{l'=0}^{\infty}
  {\cal K}_{ll'}^{I}(W,W')
  {\mbox{Im}}{\cal M}_{l}^{I}(W').
\end{equation}
The integral kernels
 \begin{equation}
   \label{Mat_kernel}
   {\cal K}_{ll'}^{I}(W,W')=\int\limits_{-1}^{1}dx
     {\mbox{\bf D}}_{l}(x){\mbox{\bf H}}(W){\mbox{\bf
         E}}^{I}(W,W',x'){\mbox{\bf H}}^{-1} (W'){\mbox{\bf
         G}}_{l'}(x')
 \end{equation}
couple the multipoles with each 
other. Analytic expressions of these kernels have been given in 
Refs.~\cite{Ber67,Han96a} for the $S$ and $P$ waves. These kernels are regular 
kinematical functions except for the diagonal kernels ${\cal K}_{ll}^{I}$, 
which contain a term $\propto 1/(W-W')$. Our further analysis is based on 
Eqs.~(\ref{multi_int}), which may be solved by several methods. In 
\cite{Ber67} conformal mapping techniques have been used. The work of 
Schwela and Weizel \cite{Sch69} follows the method of 
Omn\`es \cite{Omn58}. Lebedev 
\cite{Leb89} uses Pad\'e approximants. Since our aim is to analyse new 
experimental data and not to give mere predictions, we essentially follow 
the work of \cite{Sch69} leading to a natural parametrization of the 
multipoles. Before we turn to this point in the 
next section, we have to make some remarks.

The dispersion relations (\ref{fixed_t_photo}) are valid for the isospin 
components with $I=\pm,0$. Since the unitarity condition (\ref{multi_wat}) 
will be imposed on the solutions for the multipoles, we have to work with 
amplitudes of definite isospin $\frac{1}{2}$ or $\frac{3}{2}$. This 
leads to a mixing of the components with $I=+$ and $I=-$. The corresponding 
linear combinations of the integral kernels are obtained as follows. 
If the kernel connecting the multipoles ${\cal M}_{l'}^{I'}$
and ${\cal M}_{l}^{I}$ ($I=0,\frac{1}{2},\frac{3}{2}$ in 
the following) is denoted 
by ${\cal K}_{ll'}^{II'}$, Eqs.~(\ref{iso_iso_13}) lead to the relations
\begin{eqnarray}
  \label{iso_kernel}
  I'=\frac{3}{2} & : & \hspace{3mm}{\cal K}_{ll'}^{II'}=
    \frac{1}{3}\left(2c_{+}^{I}{\cal K}_{ll'}^{+}-c_{-}^{I}
    {\cal K}_{ll'}^{-}\right), \nonumber \\
  I'=\frac{1}{2} & : & \hspace{3mm}{\cal K}_{ll'}^{II'}=
    \frac{1}{3}\left(c_{+}^{I}{\cal K}_{ll'}^{+}+c_{-}^{I}
    {\cal K}_{ll'}^{-}\right), \\
  I'=0 & : & 
      \hspace{3mm}{\cal K}_{ll'}^{II'}=c_{0}^{I}{\cal K}_{ll'}^{0},
\nonumber
\end{eqnarray}
with $c_{+}^{3/2}=c_{+}^{1/2}=c_{0}^{0}=1$, 
$c_{-}^{3/2}=-1$, $c_{-}^{1/2}=2$ and all other coefficients 
vanishing.

In this work we restrict ourselves to the energy region 
up to $E_{\gamma}\approx$ 500 MeV. The first 
reason for this are possible problems 
with the convergence of the partial wave expansion which might occur 
in the unphysical region, when the path of the dispersion integrals leads 
outside of the Lehmann ellipses \cite{Don72}. Furthermore, in this energy 
region it is reasonable 
to restrict the solution of the integral equations to $S$, $P$ and 
eventually $D$ waves. At energies near the second resonance region, higher 
partial waves have to be taken into account which will considerably complicate 
the procedure. Finally our method provides the most reliable results 
as long as the complex phases of the partial waves are known. This 
is the case below two-pion threshold. Nevertheless it can be assumed that 
the effects of inelasticities remain small also at somewhat higher 
energies. As a consequence of these restrictions, our results will 
be especially useful for the study of low energy amplitudes and 
the electromagnetic excitation of the $\Delta(1232)$.

\section{Solution of the integral equations}

On condition that the complex 
phases of the solutions are known, Omn\`es \cite{Omn58} has proposed to 
solve the integral equations (\ref{multi_int}) as follows. Introducing 
the functions
\begin{equation}
  \label{h_func}
  h_{l}^{I}(W) = e^{i\phi_{l}^{I}(W)}\sin\phi_{l}^{I}(W),
\end{equation}
with the complex phase $\phi_{l}^{I}(W)$ of the multipole 
${\cal M}_{l}^{I}$, one obtains
\begin{equation}
  {\mbox{Im}}{\cal M}_{l}^{I}(W) 
  = h_{l}^{I\ast}(W){\cal M}_{l}^{I}(W).
\end{equation}

Next one defines the amplitudes
\begin{equation}
  \label{Multi_neu}
  {\mbox M}_{l}^{I}(W) = r_{l}(W){\cal M}_{l}^{I}
\end{equation}
by taking out kinematical factors for the $S$ and $P$ wave amplitudes,
\begin{eqnarray}
  \label{r_l}
  r_{0+} & = & \frac{W}{D_{i}D_{f}(W-m_{N})}, \nonumber \\
  r_{1-} & = & \frac{WD_{f}}{\mid\vec k\mid(W-m_{N})D_{i}}, \\
  r_{1+}(W) & = & \frac{W}{D_{i}D_{f}(W-m_{N})\mid\vec k\mid\mid\vec q\mid},
  \nonumber
\end{eqnarray}
with 
$D_{i,f} = \sqrt{E_{i,f}+m_{N}}$. The coupled system of integral equations is now cast into the form 
\begin{eqnarray}
\label{Om_coup}
{\mbox{M}}_{l}^{I}(W) & = & {\mbox{M}}_{l}^{I,\mbox{pole}}(W) +
  \frac{1}{\pi}\int\limits_{W_{\mbox{thr}}}^{\infty}\frac{h_{l}^{I\ast}(W')
  {\mbox{M}}_{l}^{I}(W')dW'}{W'-W-i\varepsilon} \nonumber \\ & & 
 +\frac{1}{\pi}\sum_{l',I'}\int\limits_{W_{\mbox{thr}}}
  ^{\infty}{\mbox{K}}_{ll'}^{II'}(W,W')h_{l'}^{I'\ast}(W')
  {\mbox{M}}_{l'}^{I'}(W')dW',
\end{eqnarray}
where the singular term has now been treated individually and 
subtracted from the integral kernel,
\begin{equation}
  \label{Kern_neu}
  {\mbox{K}}_{ll'}^{II'}(W,W')= \frac{r_{l}(W)}{r_{l}(W')}
  {\cal K}_{ll'}^{II'}(W,W')-\frac{\delta_{ll'}\delta_{II'}}{W'-W}.
\end{equation}

In order to treat Eqs.~(\ref{Om_coup}), some further simplifications 
and assumptions are necessary. First of all, the infinite sum 
over $l'$ has to be cut off. It can be shown that for multipoles 
with small complex phases in the relevant energy region, 
the solutions of the integral equations are essentially 
given by the inhomogeneities, i.~e.~the pole terms. Therefore we 
solve Eqs.~(\ref{Om_coup}) only for the $S$, $P$ and $D_{13}$ waves, 
and represent the higher partial waves by the inhomogeneities. A study of the 
integral kernels shows that many of them can be neglected. This leads to 
further simplifications.

To solve Eqs.~(\ref{Om_coup}) the complex phases $\phi_{l}^{I}(W)$ of the 
multipoles have to be known over the whole range of integration. From 
the Fermi-Watson theorem (\ref{multi_wat}) we know that these 
phases are equal to the 
pion-nucleon scattering phase shifts below two-pion threshold. 
Above the inelastic threshold, $\phi_{l}^{I}(W)$ is no longer uniquely 
determined by unitarity. However 
it has been shown that the following ansatz 
for $\phi_{l}^{I}(W)$ can be motivated from the unitarity relation of the 
inelastic amplitudes \cite{Sch69},
\begin{equation}
  \label{uni_im_2}
  \phi_{l}^{I}(W)=\arctan\left(\frac{1-\eta_{l}^{I}(W)\cos 2\delta_{l}^{I}(W)}
  {\eta_{l}^{I}(W)\sin 2\delta_{l}^{I}(W)}\right).
\end{equation}
Below two-pion threshold the ansatz (\ref{uni_im_2}) leads to 
the $\pi N$ scattering phase 
shifts $\delta_{l}^{I}$. Above this threshold the inelasticity parameters 
$\eta_{l}^{I}$ as determined from pion-nucleon scattering enter into the 
formula. The ansatz (\ref{uni_im_2}) is justified if both $\gamma N\to n\pi N$ 
and $\pi N\to n\pi N$ are dominated by the same resonant intermediate state 
or if both matrix elements are real (e.g. if they are both described by one 
particle exchange) or imaginary (e.g. in the case of purely diffractive 
processes). In the case of the $P_{13}$ waves the ansatz (\ref{uni_im_2}) would 
lead to an unphysical resonant behaviour. Therefore in this case 
we choose the alternative ansatz
\begin{equation}
  \label{uni_im_1}
  \phi_{l}^{I}(W)=\arctan\left(\frac{\eta_{l}^{I}(W)\sin 2\delta_{l}^{I}(W)}
  {1+\eta_{l}^{I}(W)\cos 2\delta_{l}^{I}(W)}\right)
\end{equation}
which has also been derived in Ref.~\cite{Sch69}.

At higher energies, where information from pion-nucleon 
scattering is no longer available,  certain assumptions have to be made 
in order to interpolate the phases to infinity \cite{Sch69}.
As an alternative we cut off the 
integrals of Eq.~(\ref{Om_coup}) at $W=2$ GeV and
represent the high energy tails of the dispersion integrals 
by $t$-channel exchange 
of $\rho$- and $\omega$-mesons, leading to the contributions:
\begin{eqnarray}
  \label{inv_vec}
  A_{1}^{\mbox{V}} & = & \frac{\lambda_{\mbox{V}}}{m_{\mbox{V}}}
     \frac{g_{\mbox{V}}^{{\cal T}}}{2m_{N}}
     \frac{t}{t-m_{\mbox{V}}^{2}},\hspace{3mm}
  A_{2}^{\mbox{V}} = -\frac{\lambda_{\mbox{V}}}{m_{\mbox{V}}}
     \frac{g_{\mbox{V}}^{{\cal T}}}{2m_{N}}
     \frac{1}{t-m_{\mbox{V}}^{2}}, \nonumber \\
  A_{3}^{\mbox{V}} & = & 0,\hspace{3mm}
  A_{4}^{\mbox{V}} = -\frac{\lambda_{\mbox{V}}}{m_{\mbox{V}}}
     g_{\mbox{V}}^{{\cal V}}
     \frac{1}{t-m_{\mbox{V}}^{2}},
\end{eqnarray}
where V stands for $\omega$ or $\rho$ respectively, 
$\lambda_{\mbox{V}}$ is taken
from the radiative decay of the vector mesons and 
$g_{\mbox{V}}^{{\cal V}, {\cal T}}$
are the vector and tensor couplings 
at the VNN vertex.
The integral equations now take the form
\begin{eqnarray}
\label{mul_int_fin}
{\mbox{M}}_{l}^{I}(W) & = & {\mbox{M}}_{l}^{I,\mbox{pole}}(W) +
  \frac{1}{\pi}\int\limits_{W_{\mbox{thr}}}^{\Lambda}\frac{h_{l}^{I\ast}(W')
  {\mbox{M}}_{l}^{I}(W')dW'}{W'-W-i\varepsilon} 
 \\ & &
 + \frac{1}{\pi}\sum_{l',I'}\int\limits_{W_{\mbox{thr}}}
  ^{\Lambda}{\mbox{K}}_{ll'}^{II'}(W,W')h_{l'}^{I'\ast}(W')
  {\mbox{M}}_{l'}^{I'}(W')dW'
  + {\mbox{M}}_{l}^{I,\mbox{V}}(W), \nonumber
\end{eqnarray}
with a cut-off at $\Lambda=2$ GeV. The high energy parametrization by 
$t$-channel exchange leads to a further 
inhomogeneity of the integral equations, ${\mbox{M}}_{l}^{I,\mbox{V}}(W)$, 
to be constructed according to the rules given above. In comparison with 
\cite{Omn58} this leads to a slight modification
of the solutions. As in \cite{Omn58} we first solve the characteristic
equations
\begin{eqnarray}
\label{mul_char}
{\mbox{M}}_{l}^{I}(W) & = & {\mbox{M}}_{l}^{I,\mbox{pole}}(W) +
  \frac{1}{\pi} \int\limits_{W_{\mbox{thr}}}^{\Lambda}\frac{h_{l}^{I\ast}(W')
  {\mbox{M}}_{l}^{I}(W')dW'}{W'-W-i\varepsilon} \nonumber \\ & & 
  + {\mbox{M}}_{l}^{I,\mbox{V}}(W),
\end{eqnarray}
which contain only the singular part of the integral equations. The general 
solutions are
\begin{eqnarray}
\label{los_char}
{\mbox{M}}_{l}^{I}(W) & = & {\mbox{M}}_{l}^{I,\mbox{pole}}(W) +
   {\mbox{M}}_{l}^{I,\mbox{V}}(W) \nonumber \\ & & +
   \frac{{\mbox{M}}_{l,0}^{I,\mbox{hom}}(W)}{\pi}
   \int\limits_{W_{\mbox{thr}}}^{\Lambda}\frac{h_{l}^{I}(W')
  ({\mbox M}_{l}^{I,\mbox{pole}}(W')
  +{\mbox{M}}_{l}^{I,\mbox{V}}(W'))dW'}
  {{\mbox{M}}_{l,0}^{I,\mbox{hom}}(W')(W'-W-i\varepsilon)} \nonumber \\ & & 
  +P(W){\mbox{M}}_{l,0}^{I,\mbox{hom}}(W),
\end{eqnarray}
with
\begin{equation}
  \label{sol_hom}
 {\mbox{M}}_{l,0}^{I,\mbox{hom}}(W) = \exp\left\{\frac{W}{\pi}
 \int\limits_{W_{\mbox{thr}}}^{\Lambda}\frac{\phi_{l}^{I}(W')dW'}
  {W'(W'-W-i\varepsilon)} \right\}
\end{equation}
a solution of the homogeneous equation
\begin{equation}
\label{char_hom_eq}
{\mbox{M}}_{l}^{I}(W) =
  \frac{1}{\pi}\int\limits_{W_{\mbox{thr}}}^{\Lambda}\frac{h_{l}^{I\ast}(W')
  {\mbox{M}}_{l}^{I}(W')dW'}{W'-W-i\varepsilon},
\end{equation}
and $P(W)$ an arbitrary polynomial. In our calculations for this polynomial 
only the simplest case, i.e. a real constant, will occur.

If the self-coupling of ${\mbox{M}}_{l}^{I}(W)$ is fully taken into account,
Eq. (\ref{mul_int_fin}) has to be transformed into an equation of the
Fredholm type,
\begin{equation}
  \label{fred_mul}
  {\mbox{M}}_{l}^{I}(W)=N_{l}^{I}(W)+\sum_{l',I'}
  \frac{1}{\pi} \int\limits_{W_{\mbox{thr}}}^{\Lambda}{\mbox{F}}_{ll'}^{II'}
   (W,W')h_{l}^{I'\ast}(W'){\mbox{M}}_{l'}^{I'}(W')dW',
\end{equation}
with
\begin{eqnarray}
  \label{inh_fred}
  N_{l}^{I}(W) & = & {\mbox{M}}_{l}^{I,\mbox{pole}}(W) +
   {\mbox{M}}_{l}^{I,\mbox{V}}(W)
   \\ & & + \frac{{\mbox{M}}_{l,0}^{I,{\mbox{hom}}(W)}}{\pi} 
   \int\limits_{W_{\mbox{thr}}}^{\Lambda}\frac{h_{l}^{I}(W')
  {(\mbox{M}}_{l}^{I,\mbox{pole}}(W')
  +{\mbox{M}}_{l}^{I,\mbox{V}}(W'))dW'}
  {{\mbox{M}}_{l,0}^{I,\mbox{hom}}(W')(W'-W-i\varepsilon)} \nonumber
\end{eqnarray}
and the Fredholm kernels
\begin{eqnarray}
  \label{kern_fred}
  F_{ll'}^{II'}(W,W') & = & {\mbox K}_{ll'}^{II'}(W,W')h_{l}^{I'\ast}(W')
   + \frac{{\mbox M}_{l,0}^{I,\mbox{hom}}(W)}{\pi} \nonumber \\ & &
   \times\int\limits_{W_{\mbox{thr}}}^{\Lambda}\frac{h_{l}^{I'}(W'')
   {\mbox K}_{ll'}^{II'}(W'',W')h_{l}^{I\ast}(W')dW''}
  {{\mbox M}_{l,0}^{I,\mbox{hom}}(W'')(W''-W-i\varepsilon)}.
\end{eqnarray}
Eq.~(\ref{fred_mul}) can now be solved by standard numerical methods. We 
mention that also the solution of Eq.~(\ref{fred_mul}) is not unique. A 
solution of the corresponding homogeneous equation can be added 
in the same way as in the 
case of Eq.~(\ref{mul_char}. For the mathematical details of solving the 
homogeneous equation see Ref.~\cite{Han96a}.

Numerically we found that the self-couplings of the multipoles 
are given to good approximation by the singular parts of the integral 
equations, except for the $P_{33}$ multipoles $M_{1+}^{(\frac{3}{2})}$ 
and $E_{1+}^{(\frac{3}{2})}$. Therefore, the 
Fredholm equations have to be solved for these $P$-waves, while in 
all other cases it is quite sufficient to solve the 
characteristic equations. If the couplings to other multipoles are taken into 
account, they enter into the equations as further inhomogeneities. 
If we neglect all integral kernels giving only small contributions, 
we obtain the following procedure to solve the integral equations 
for the $S$-, $P$- and $D_{13}$-waves:
\begin{enumerate}
\item The amplitude $M_{1+}^{(\frac{3}{2})}$ is practically independent of 
the other multipoles. The corresponding 
Fredholm equation has to be solved.
\item Having solved the equation for $M_{1+}^{(\frac{3}{2})}$, we calculate 
its contribution to $E_{1+}^{(\frac{3}{2})}$ entering into the 
corresponding equation as a further inhomogeneity.
\item The equations for $M_{1-}^{(\frac{1}{2})}$ and 
$M_{1-}^{(\frac{3}{2})}$ are
solved taking account of the influence of $M_{1+}^{(\frac{3}{2})}$ and
$E_{1+}^{(\frac{3}{2})}$.
\item For $E_{0+}^{(\frac{1}{2})}$ and $E_{0+}^{(\frac{3}{2})}$ also the 
contributions of the $M_{1-}$ components are important.
\item In the isospin 0 channel, only the contribution of $M_{1-}^{(0)}$ to 
$E_{0+}^{(0)}$ has to be taken into account.
\item For all other $P$ waves as well as the $D_{13}$ multipoles we 
only have to solve the characteristic equations.
\end{enumerate}

According to Eq.~(\ref{los_char}) we can add to each solution 
of the inhomogeneous equation 
a solution of the homogeneous equation multiplied by an arbitrary 
polynomial in $W$. Following the arguments of \cite{Sch69}, we restrict 
ourselves to the addition of homogeneous solutions to the multipoles
$E_{0+}^{(0)}$, 
$E_{0+}^{(\frac{1}{2})}$, $M_{1-}^{(0)}$, $M_{1-}^{(\frac{1}{2})}$, 
$E_{1+}^{(\frac{3}{2})}$ and $M_{1+}^{(\frac{3}{2})}$ multiplied by a 
real constant. These six constants are determined by a fit to the data.
Further parameters of our fit are the coupling constants 
of the vector mesons. 
In \cite{Sch69} it is argued that vector meson exchange should not be 
included because of double counting. This argument is 
correct because the authors evaluate the occuring integrals to infinity. From 
a study of unitarity diagrams it can be seen that the integrals in fact 
contain the vector meson contributions. In our case, however, the integrals 
run only to the cut off $\Lambda$, and only a fraction 
of vector meson exchange will be contained in the integrals. 
The remaining part is assumed to be the dominant contribution at the higher 
energies, which we represent by vector meson exchange as described 
with four free VNN coupling constants fitted to the data. 
Thus we end up with ten free parameters. 

In solving the integral equations we have to know the scattering phase 
shifts and inelasticity parameters of pion nucleon scattering. They are 
taken from the SAID code (solution SM95 \cite{Arn95}).

\section{The fit to the data}
We have determined the ten free parameters of our approach 
by a fit to selected 
photoproduction data for 160 MeV $\leq$ $E_{\gamma}$ $\leq$ 420 MeV 
(see Tab.~\ref{tab:data}). Most 
interesting for our analysis are the new data from MAMI. Differential 
cross sections of $\pi^{0}$ production off the proton near threshold 
\cite{Fuc96}, and 
differential cross sections as well as beam asymmetries $\Sigma$ for 
both $\pi^{+}$ and $\pi^{0}$ production off the proton 
\cite{Kra96,Hae96} have been 
measured with high precision. Since there are no new low energy data 
for $\pi^{+}$ we include those of the data 
compilation \cite{Men77}. We have also included 
the most recent data from ELSA, 
differential cross sections for $\pi^{+}$ 
production \cite{Bue94} and the target asymmetries $T$ for $\pi^{+}$ and 
$\pi^{0}$ production off the proton \cite{Dut95}. To get the full isospin 
decomposition we also had to include data for $\pi^{-}$ production, for 
which we took the differential cross sections given in 
Refs.~\cite{Ben58,Car73,Ben73,Fuj76,Bag88}. Altogether we have included 
1751 data points. The overall $\chi^{2}$ per data point of 2.8. We note that 
we have not renormalized the different data sets in our fit.

\begin{figure}[htbp]
\centerline{\psfig{figure=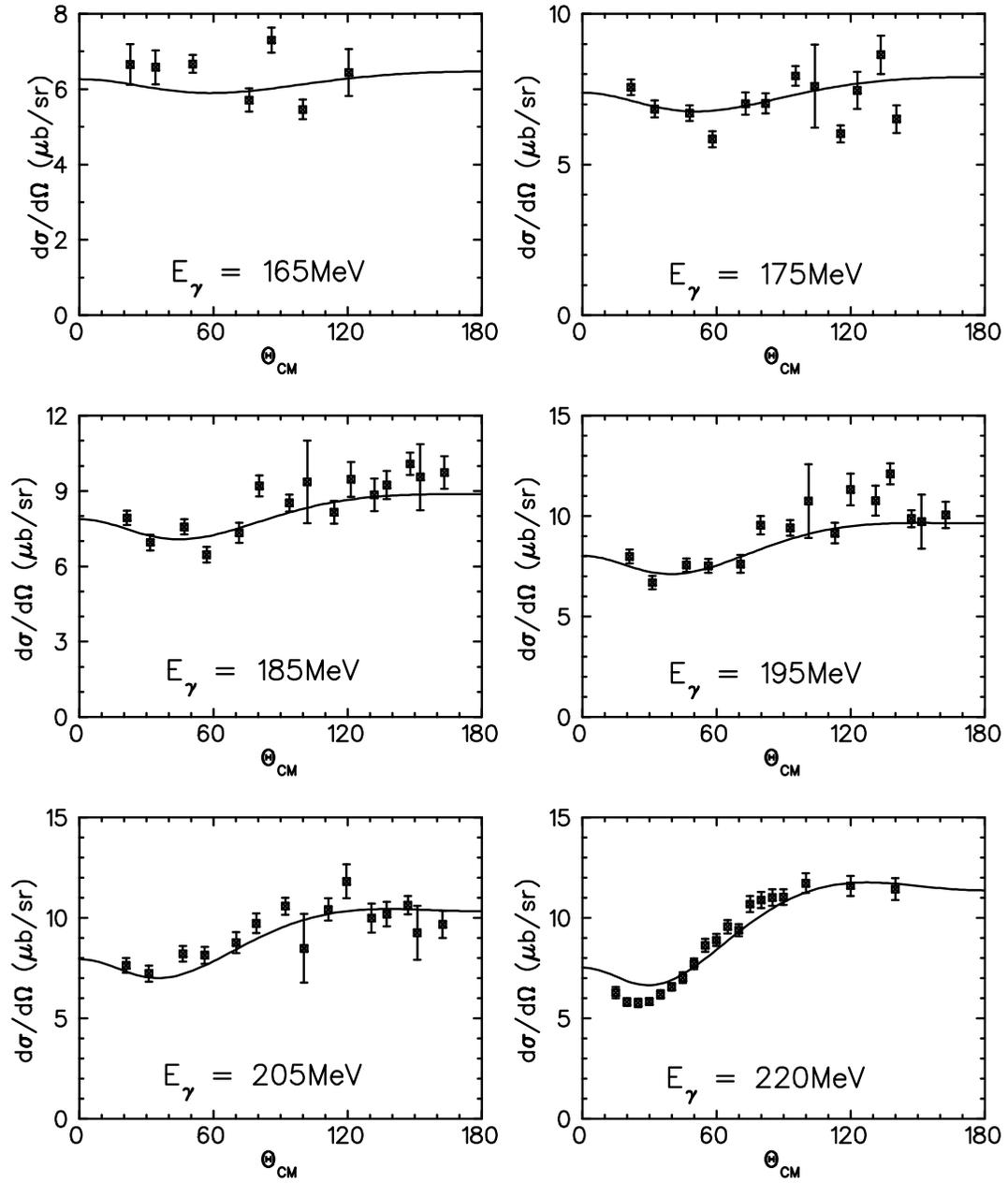,width=14cm}}
    \caption{Differential cross sections for $\gamma p\to\pi^{+}n$ at 
         lower energies. The data are taken from Ref.~\protect\cite{Men77}.}
  \label{fig:1}
\end{figure}

\begin{figure}[htbp]
\centerline{\psfig{figure=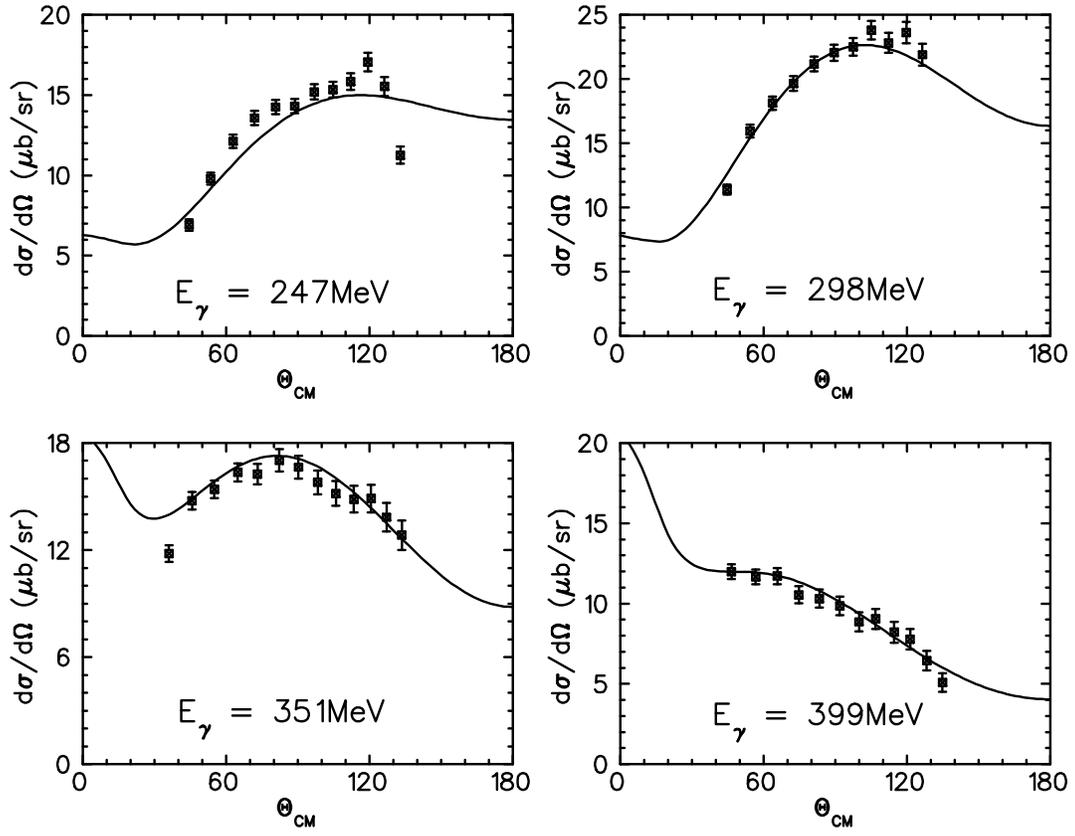,width=14cm}}
    \caption{Differential cross sections for $\gamma p\to\pi^{+}n$ 
    around the $\Delta(1232)$-resonance. The data are  
    from Bonn \protect\cite{Bue94}.}
  \label{fig:2}
\end{figure}

\begin{figure}[htbp]
\centerline{\psfig{figure=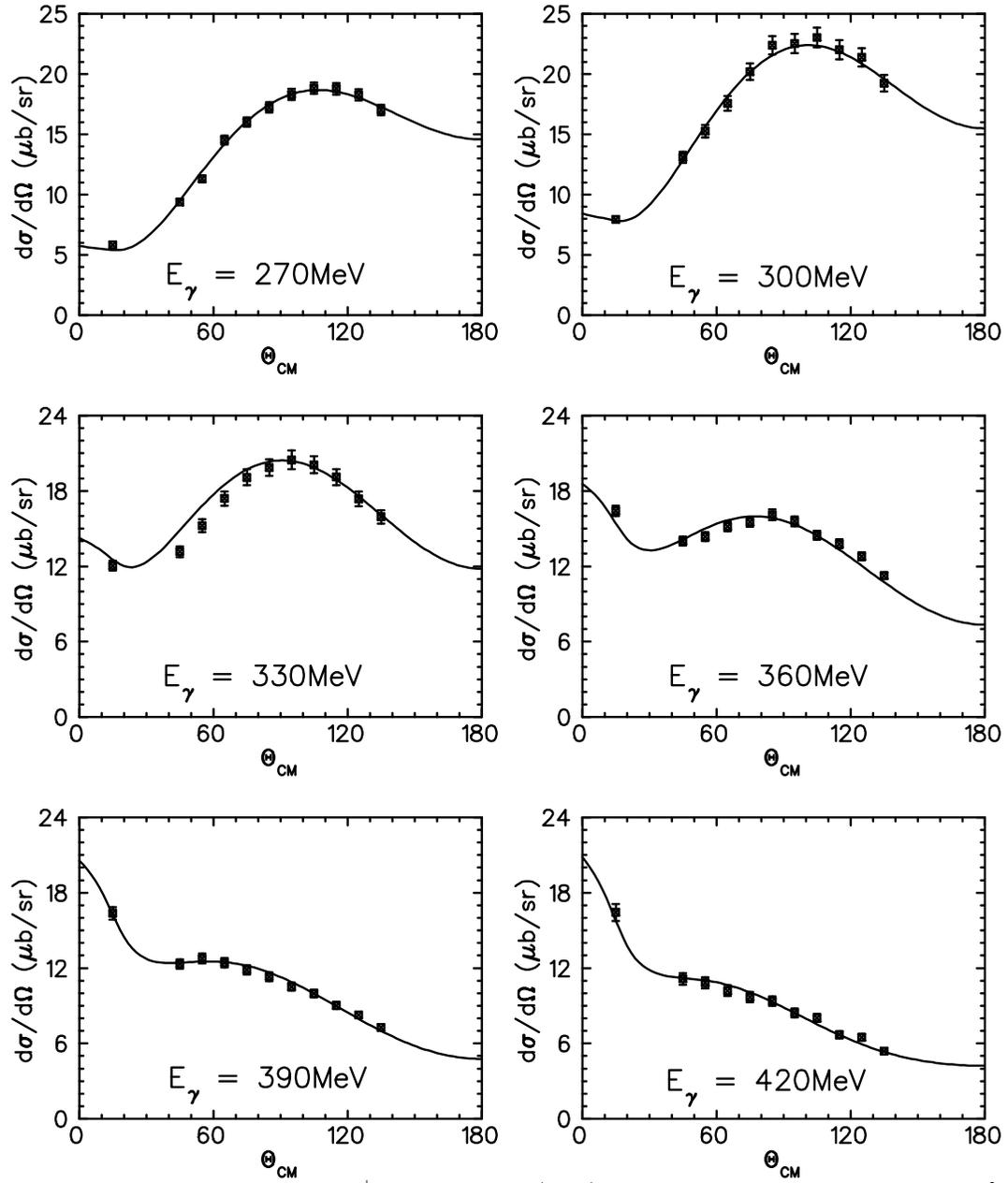,width=14cm}}
    \caption{Differential cross sections for $\gamma p\to\pi^{+}n$ 
    around the $\Delta(1232)$-resonance. The data are 
    from Mainz \protect\cite{Kra96} except the data for $\theta = 15^{\circ}$
    which are from Bonn \protect\cite{Fis72}.}
  \label{fig:3}
\end{figure}

\begin{figure}[htbp]
\centerline{\psfig{figure=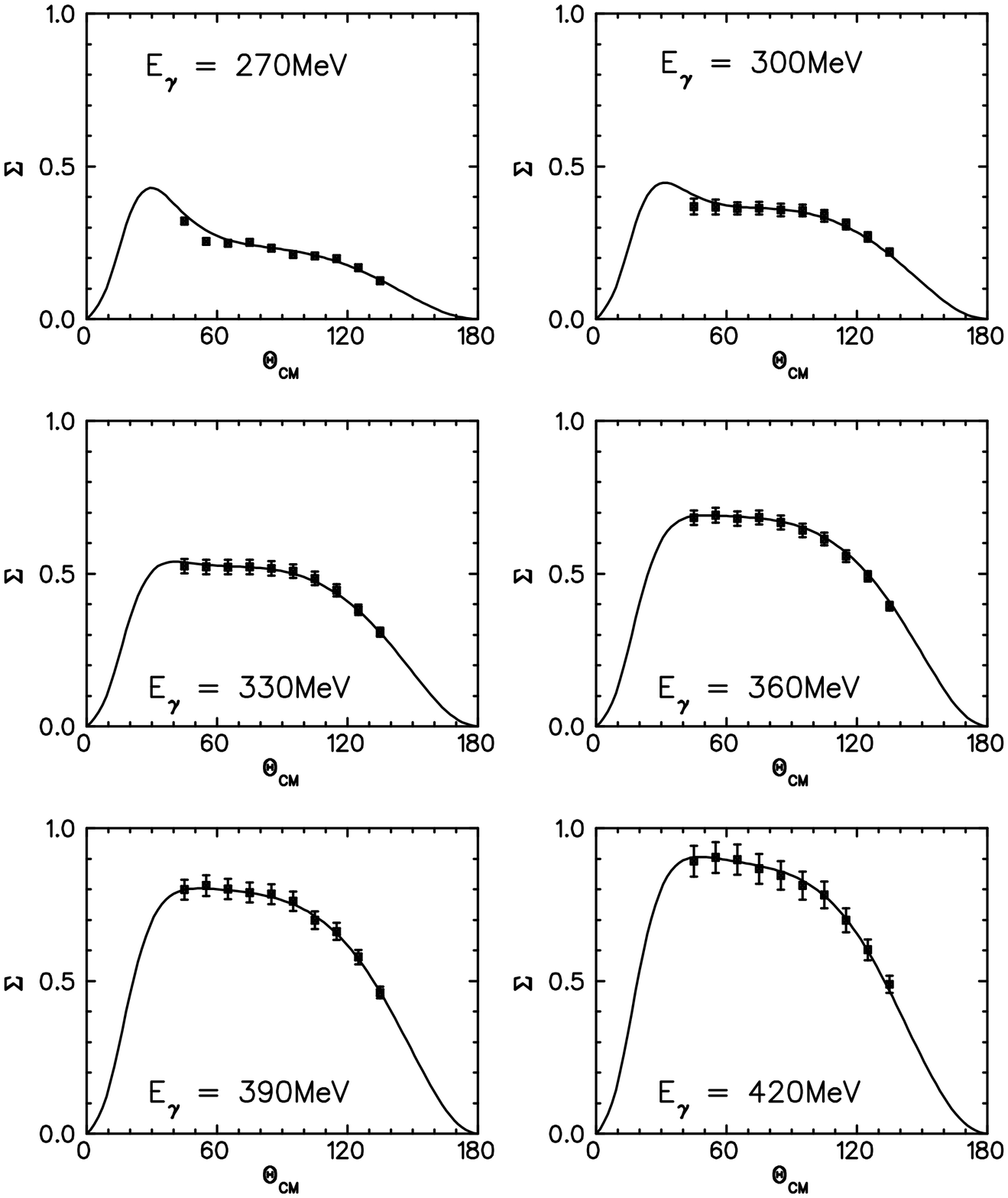,width=14cm}}
    \caption{The beam asymmetry $\Sigma$ for $\gamma p\to\pi^{+}n$. 
    The data are from  Mainz \protect\cite{Kra96}.}
  \label{fig:4}
\end{figure}

\begin{figure}[htbp]
\centerline{\psfig{figure=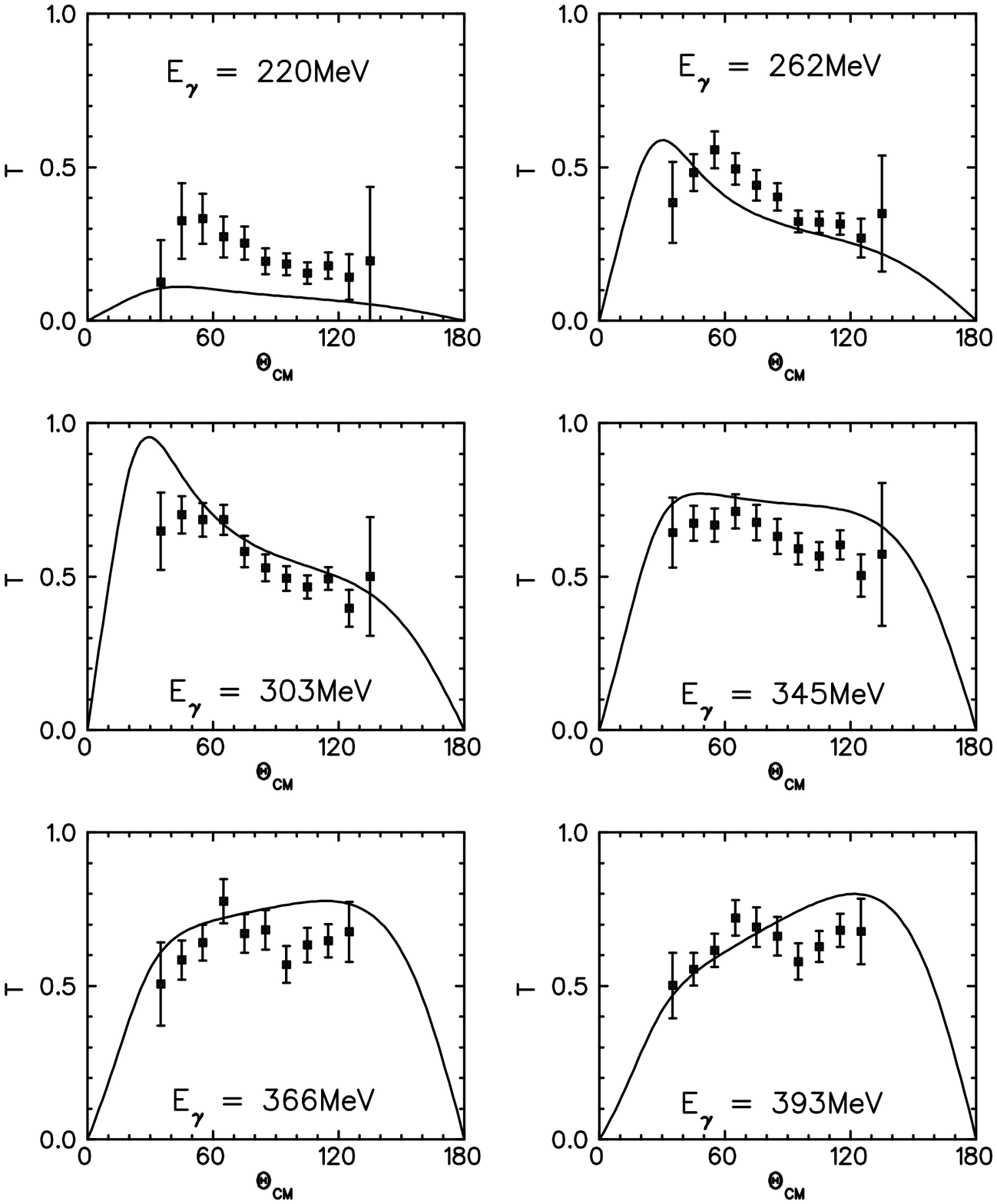,width=14cm}}
    \caption{The target asymmetry $T$ for $\gamma p\to\pi^{+}n$. 
    The data are from Bonn \protect\cite{Dut95}.}
  \label{fig:5}
\end{figure}

\begin{figure}[htbp]
\centerline{\psfig{figure=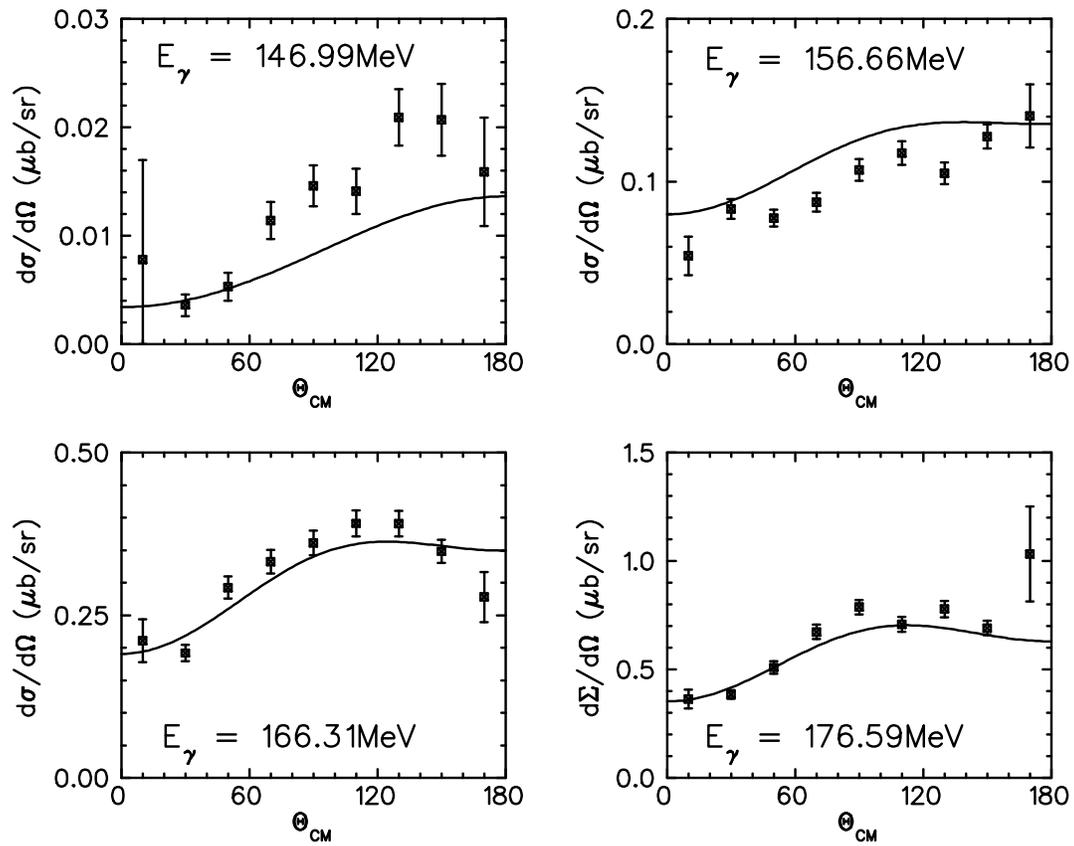,width=14cm}}
    \caption{Differential cross sections for $\gamma p\to\pi^{0}p$
    at low energies. 
    The data are from Mainz \protect\cite{Fuc96}.}
  \label{fig:6}
\end{figure}

\begin{figure}[htbp]
\centerline{\psfig{figure=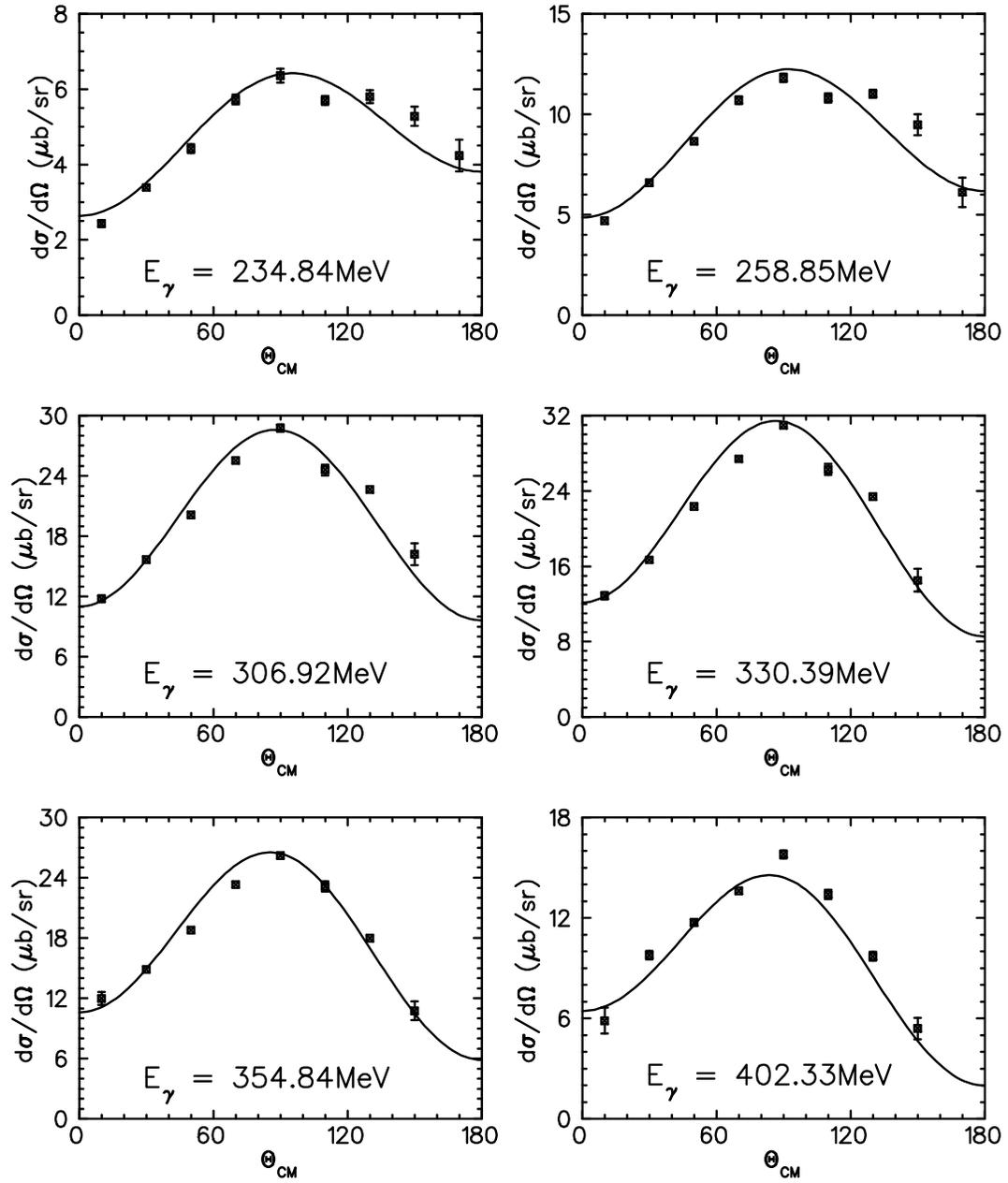,width=14cm}}
    \caption{Differential cross sections for $\gamma p\to\pi^{0}p$. 
    The data are from Mainz \protect\cite{Hae96}.}
  \label{fig:7}
\end{figure}

\begin{figure}[htbp]
\centerline{\psfig{figure=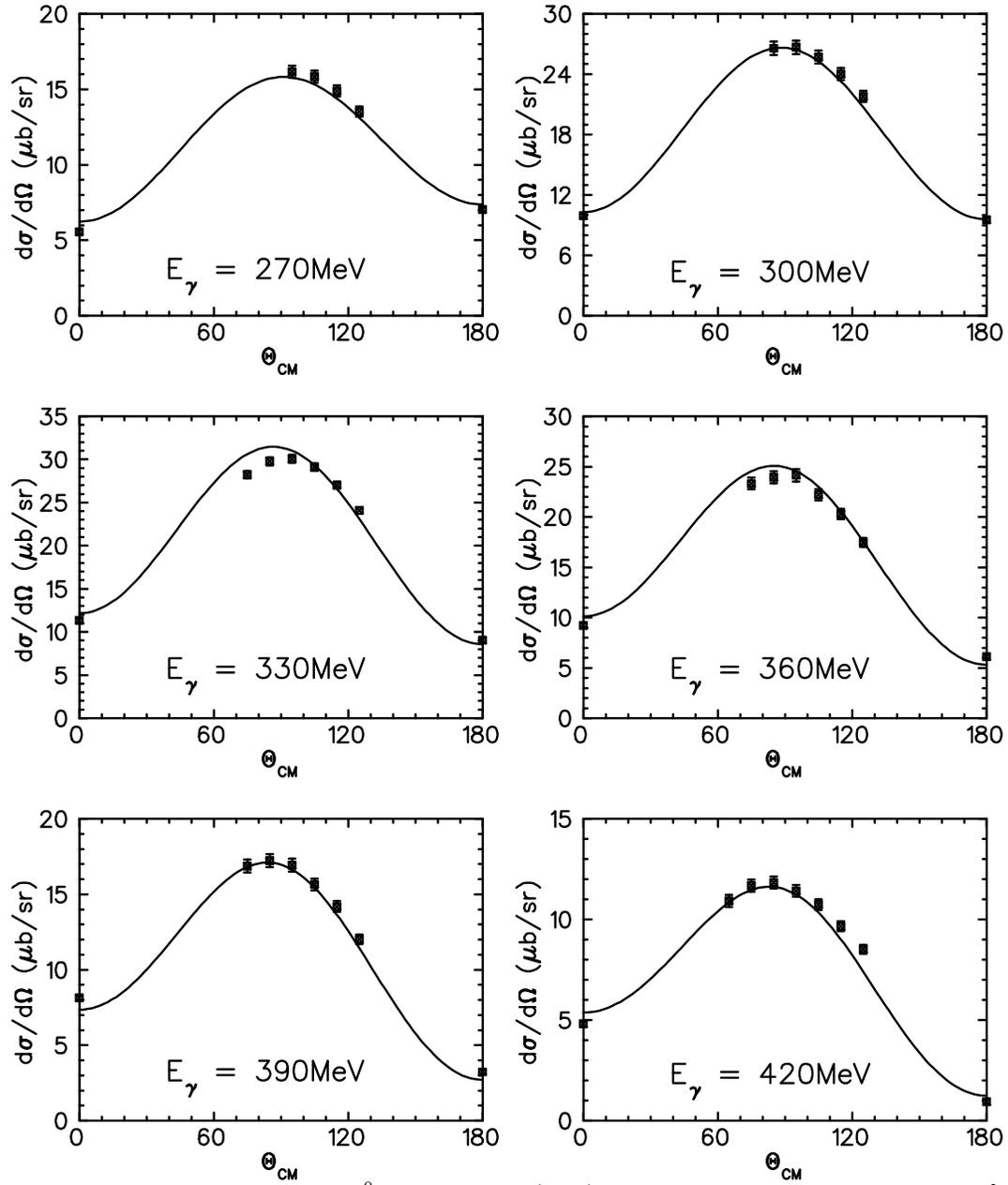,width=14cm}}
    \caption{Differential cross sections for $\gamma p\to\pi^{0}p$
     around the $\Delta(1232)$-resonance. 
    The data are from Mainz \protect\cite{Kra96} except the data 
    at the angles of $0^{\circ}$ and $180^{\circ}$ which are from
    Bonn \protect\cite{Gen74}.}
  \label{fig:8}
\end{figure}

\begin{figure}[htbp]
\centerline{\psfig{figure=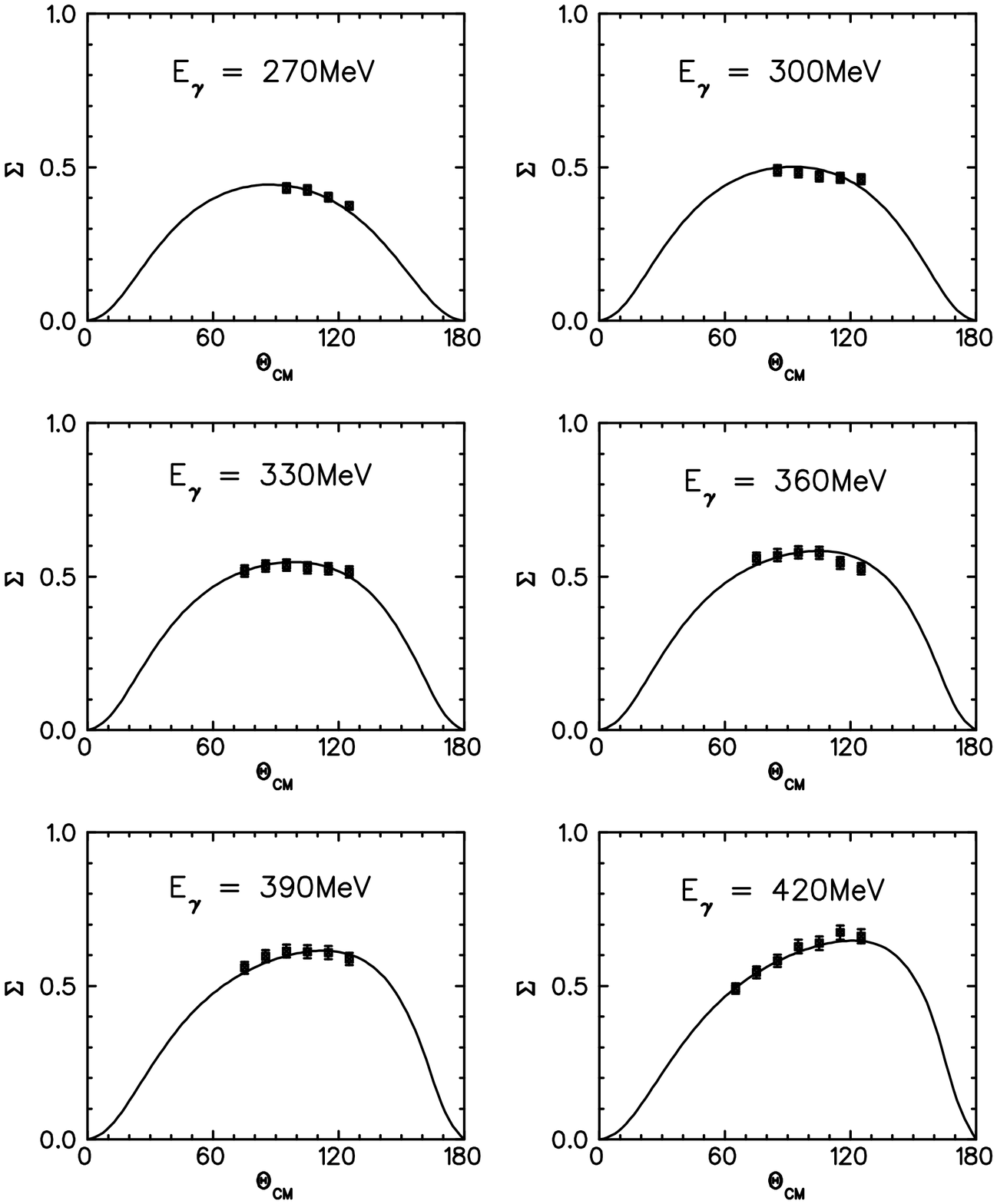,width=14cm}}
    \caption{The beam asymmetry $\Sigma$ for $\gamma p\to\pi^{0}p$. 
    The data are from Mainz \protect\cite{Kra96}.}
  \label{fig:9}
\end{figure}

\begin{figure}[htbp]
\centerline{\psfig{figure=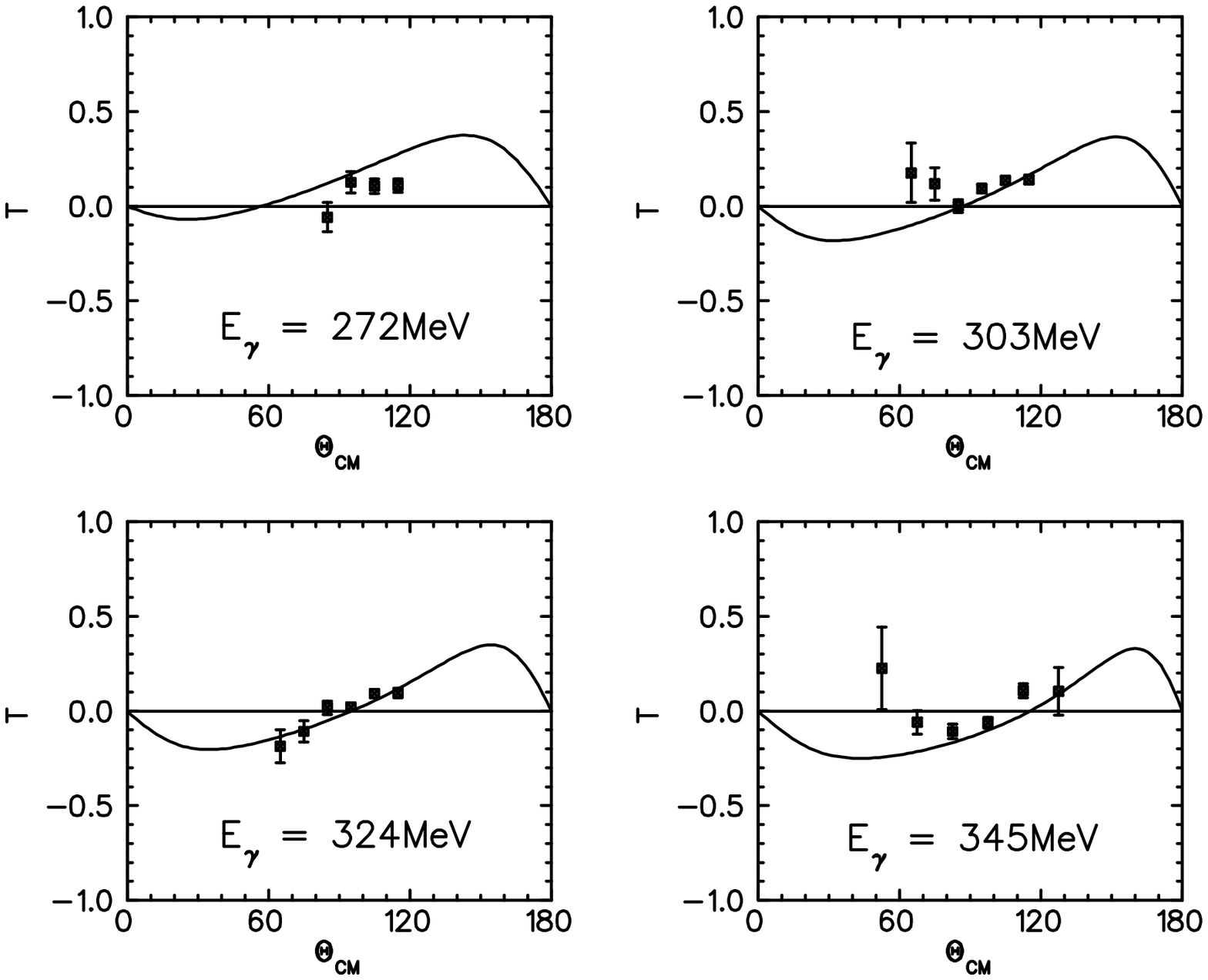,width=14cm}}
    \caption{The target asymmetry $T$ for $\gamma p\to\pi^{0}p$. 
    The data are from Bonn \protect\cite{Dut95}.}
  \label{fig:10}
\end{figure}

\begin{figure}[htbp]
\centerline{\psfig{figure=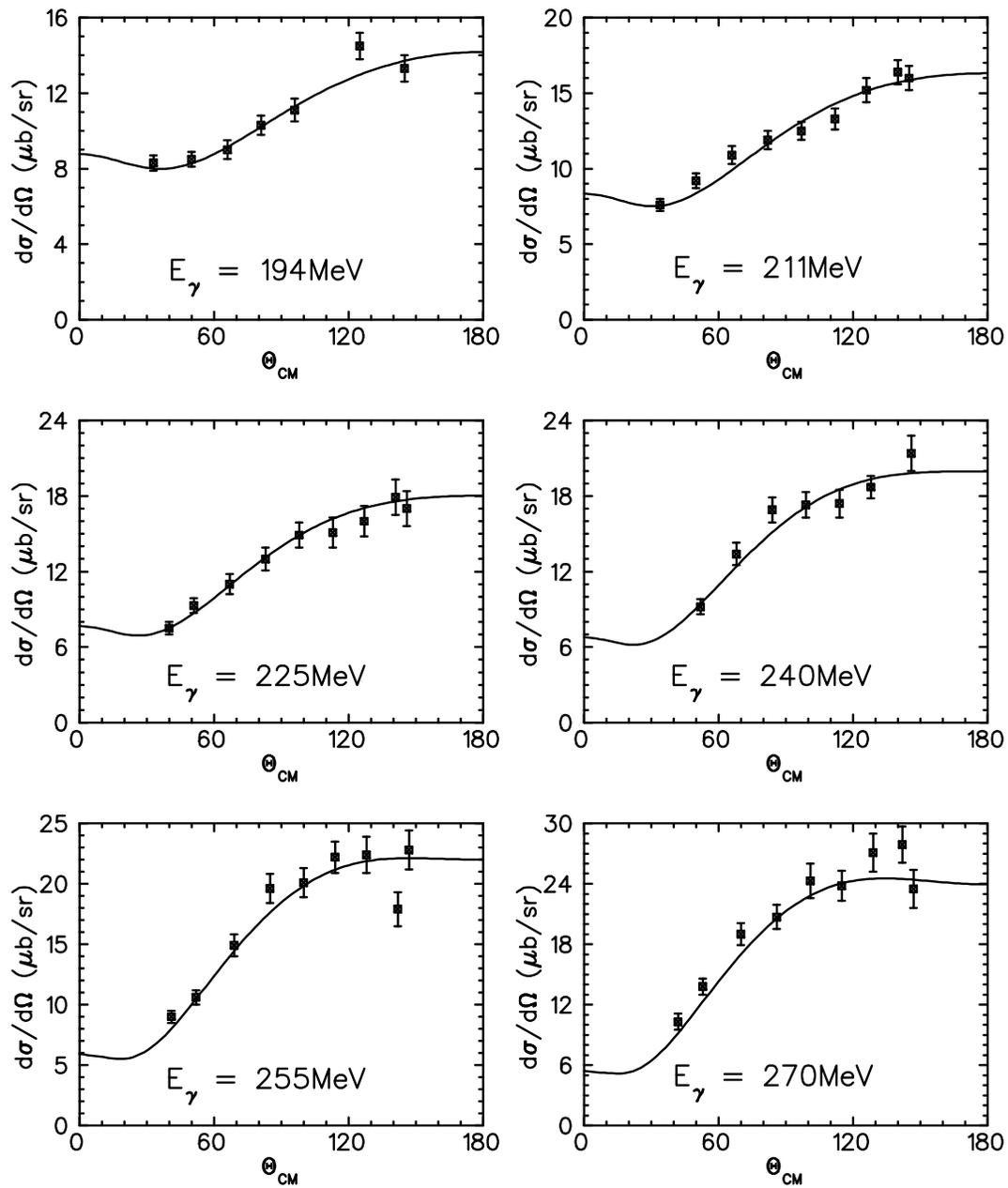,width=14cm}}
    \caption{Differential cross sections for $\gamma n\to\pi^{-}p$. 
    The data are from TRIUMF \protect\cite{Bag88}.}
  \label{fig:11}
\end{figure}

\begin{figure}[htbp]
\centerline{\psfig{figure=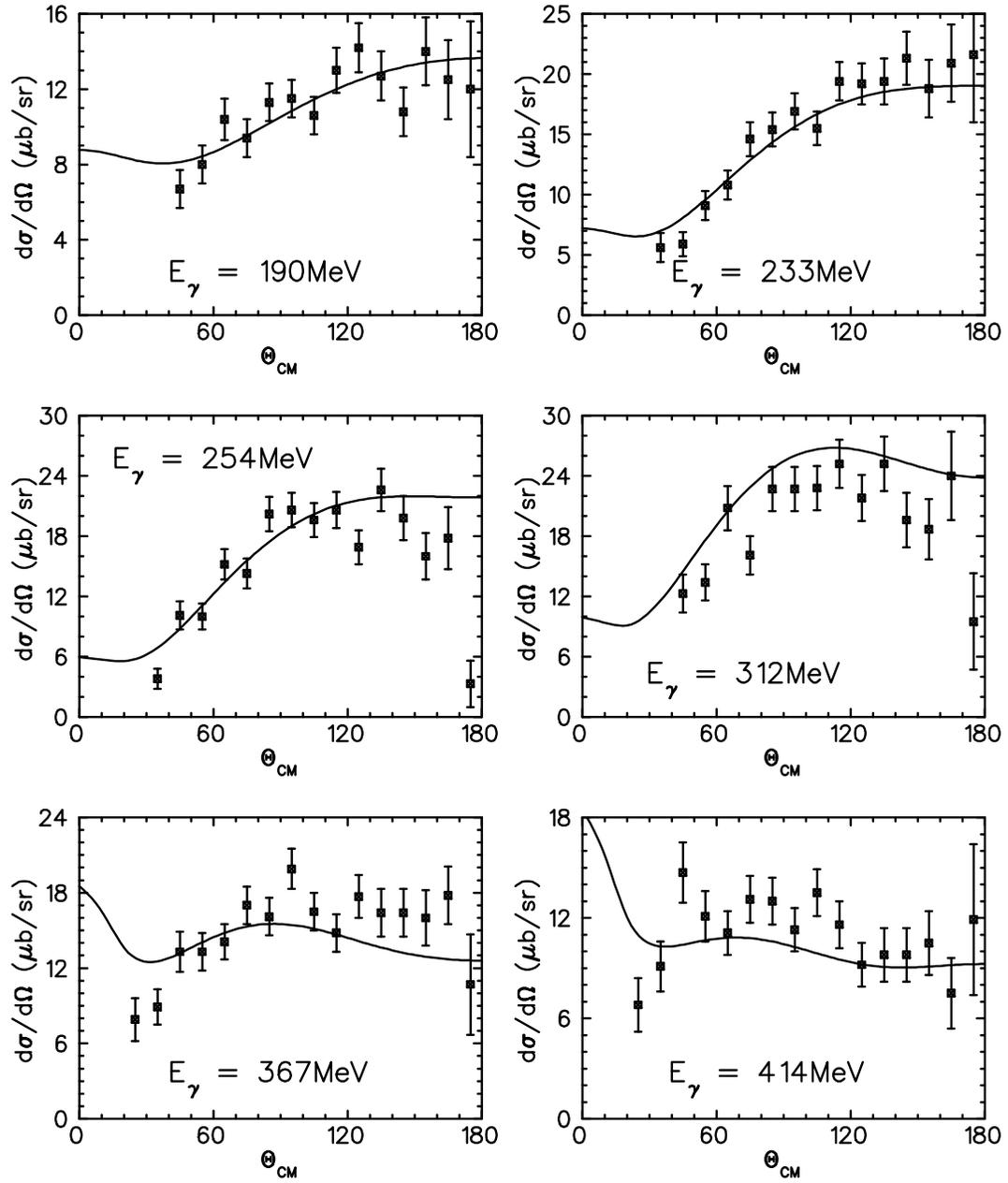,width=14cm}}
    \caption{Differential cross sections for $\gamma n\to\pi^{-}p$. 
    The data are from Frascati \protect\cite{Car73}.}
  \label{fig:12}
\end{figure}

In Figs.~\ref{fig:1}--\ref{fig:12} we show our results for the observables. 
In general we obtain 
a good description of the photoproduction data. The differential cross 
sections and beam asymmetries for $\pi^{+}$ production are in perfect 
agreement with the Mainz experiment \cite{Kra96} except for the differential 
cross sections around 330 MeV where we find a slight discrepancy 
(see Fig.~\ref{fig:3}). In our local fit (see next section) 
this discrepancy could not be removed which means that there is a 
problem either with our parametrization or with the experiment. We stress that 
also at these energies the beam asymmetries, which are less sensitive 
to systematical errors, are well described by our fit 
(see Fig.~\ref{fig:4}). From the experimental 
point of view one main difficulty are measurements concerning 
$\pi^{+}$ production at very low energies. Because of the short life time 
of charged pions many of them decay before detection. As a consequence 
the existing data have large error bars. In Fig.~\ref{fig:1} we show the 
result of the fit compared to the world data. We find good agreement 
within the statistical uncertainties. The differential cross sections and 
target asymmetries from Bonn which we included in our fit 
(see Figs.~\ref{fig:2} and \ref{fig:5}) have larger errors 
than the new Mainz data. In the case 
of the target asymmetries our result seems to be consistent with the 
data if the statistical uncertainty is taken into account, except for very 
low energies where the experimental values are systematically larger than 
our curves. However, it can be shown that such large values of $T$ would 
require much larger imaginary parts of the partial waves than allowed by 
unitarity considerations.

In $\pi^{0}$ production on the proton we find a somewhat different 
situation. In this case the results of recent high precision experiments at low 
energies are available \cite{Fuc96}. However, only the differential 
cross sections for $E_{\gamma} \geq 160$ MeV were included 
in our fit. The description is reasonably 
good also for the lower energies 
(see Fig.~\ref{fig:6}). At higher energies we again find a very nice 
description of the beam asymmetries (see Fig.~\ref{fig:9}). Also the 
differential cross sections are well described except for the energies 
around 330 MeV (see Figs.~\ref{fig:7} and \ref{fig:8}). The target 
asymmetries are only in semiquantitative agreement (see Fig.~\ref{fig:10}).

In pion production on the neutron the experimental situation is less 
satisfactory than for the proton. The most precise experiment which has 
been performed at TRIUMF by studying the inverse reaction 
$\pi^{-}p\to\gamma n$ \cite{Bag88}, is restricted to energies below 
the $\Delta$ resonance. In Fig.~\ref{fig:11} it can be seen that we fit this 
experiment very well. However, to take account of a larger energy range, we
had to include older data which are reasonably well described 
within the larger statistical uncertainties (see Fig.~\ref{fig:12}).

The parameters determined by our fit have no direct physical 
interpretation. 
Given the integral equations, the coefficients of the homogeneous 
solutions have to be determined by the 
boundary conditions determined by the physics. Concerning the 
coupling constants of 
the vector mesons, we expect them to be of the same order 
of magnitude as the corresponding values of other analyses. In 
Tab.~\ref{tab:1} we compare our results with other references. The existing 
differences should be attributed to the fact part of the vector meson 
strength is already contained in the dispersive integral up to 2 GeV.

\section{Results, predictions and discussion}

\subsection{$S$ and $P$ wave multipoles}
\begin{figure}[htbp]
\centerline{\psfig{figure=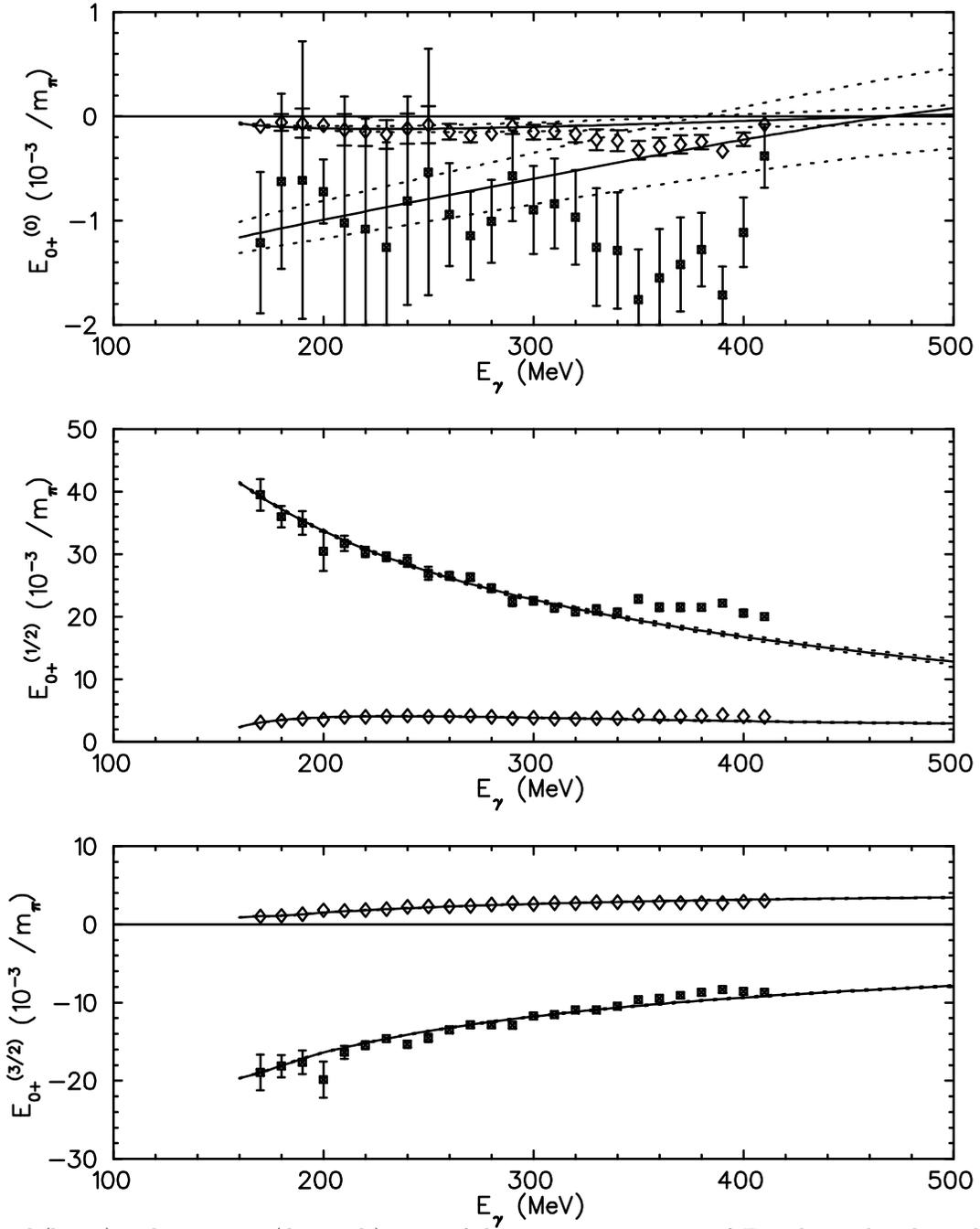,width=14cm}}
    \caption{Real (boxes) and imaginary (diamonds) parts of the 
     isospin components of $E_{0+}$ obtained with our local fit. The 
     solid lines 
     are the corresponding results of the global fit, the dashed lines 
     indicate the error band.}
  \label{fig:13}
\end{figure}

\begin{figure}[htbp]
\centerline{\psfig{figure=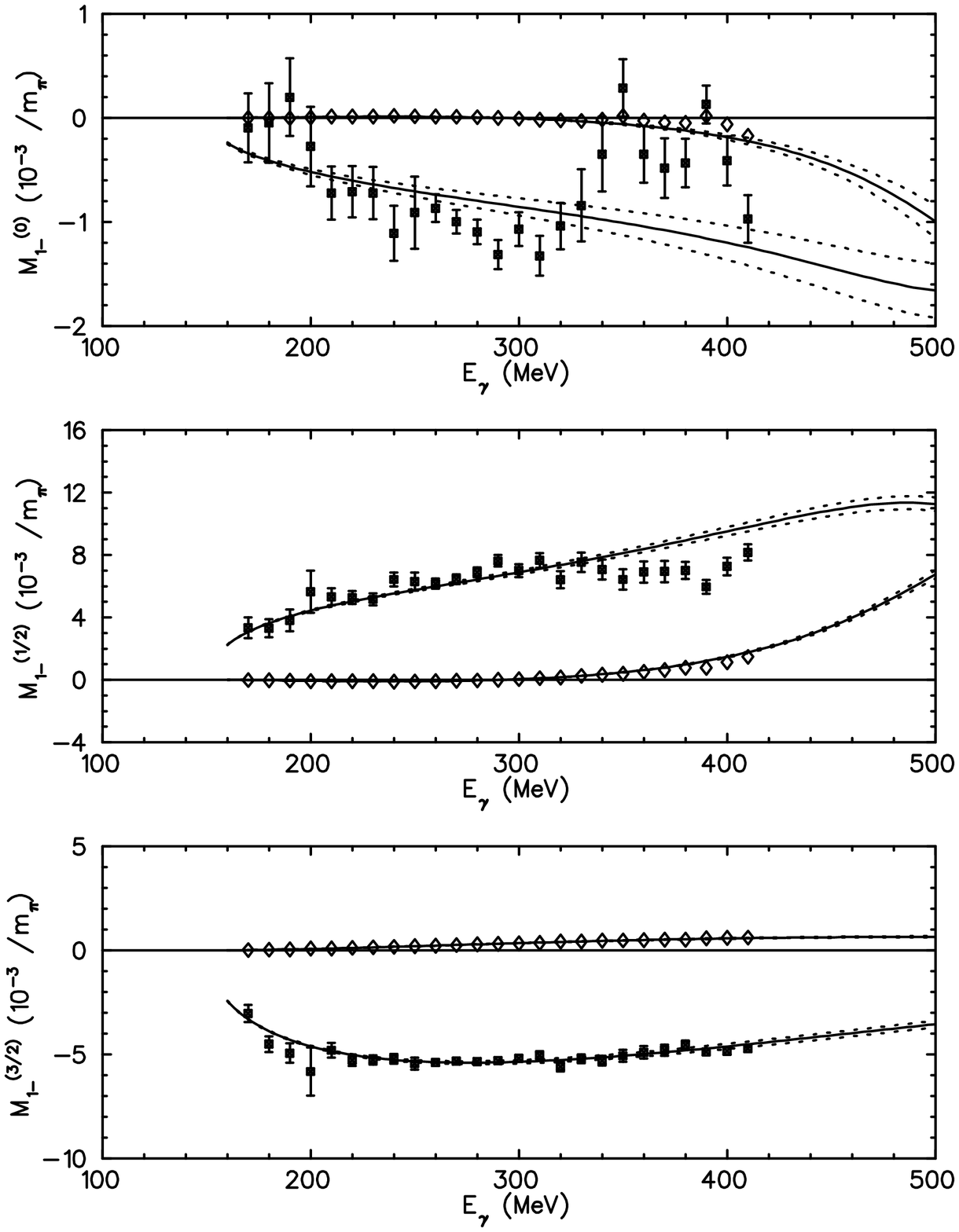,width=14cm}}
    \caption{Isospin components of $M_{1-}$. Symbols as 
             in Fig.~\protect\ref{fig:13}.}
  \label{fig:14}
\end{figure}

\begin{figure}[htbp]
\centerline{\psfig{figure=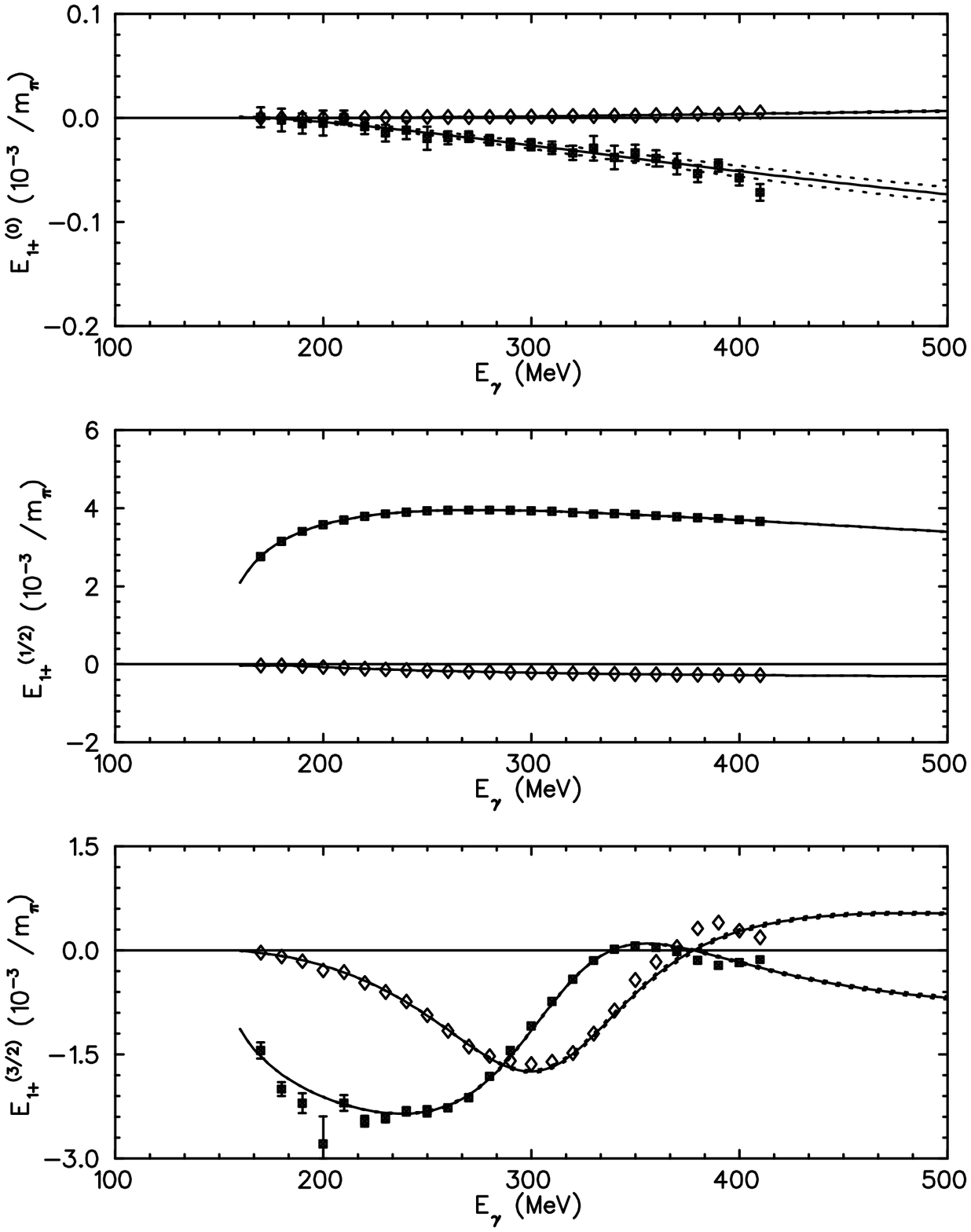,width=14cm}}
    \caption{Isospin components of $E_{1+}$. Symbols as 
             in Fig.~\protect\ref{fig:13}.}
  \label{fig:15}
\end{figure}

\begin{figure}[htbp]
\centerline{\psfig{figure=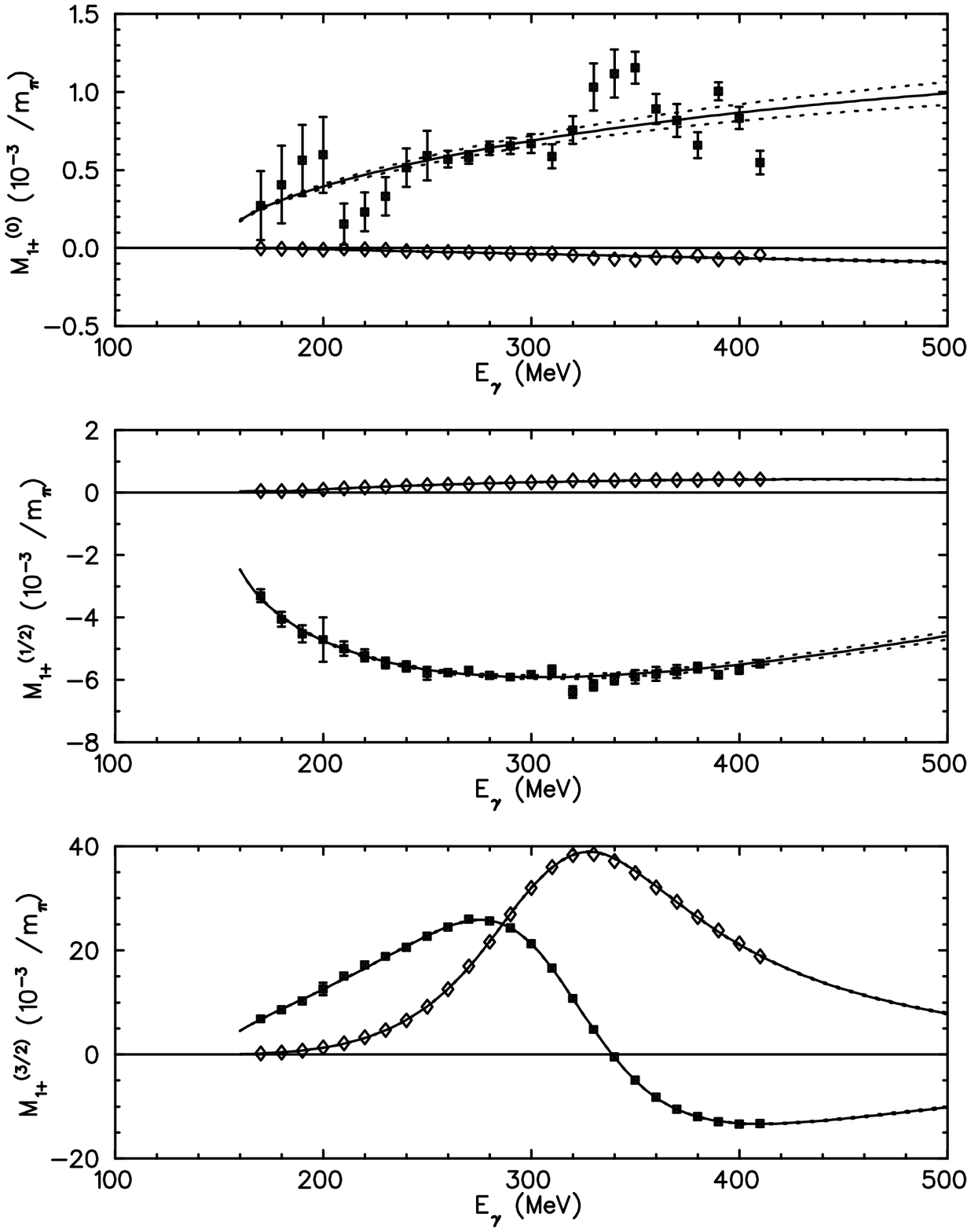,width=14cm}}
    \caption{Isospin components of $M_{1+}$. Symbols as 
             in Fig.~\protect\ref{fig:13}.}
  \label{fig:16}
\end{figure}

\begin{figure}[htbp]
\centerline{\psfig{figure=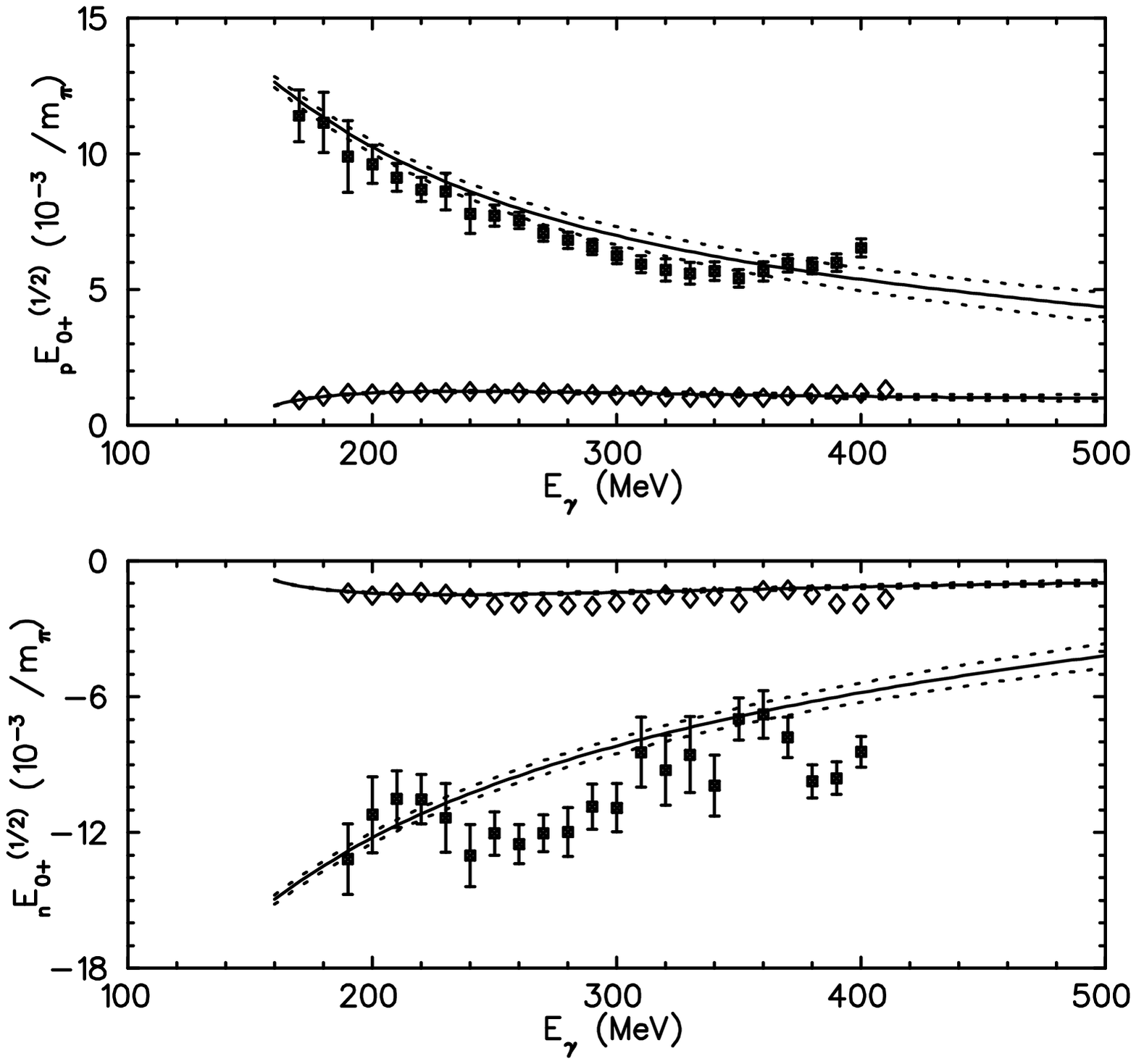,width=14cm}}
    \caption{Decomposition of the isospin $\frac{1}{2}$ channel of the
     amplitude $E_{0+}$ into components for reactions off the proton 
     and the neutron respectively. Symbols as in Fig.~\protect\ref{fig:13}.}
  \label{fig:17}
\end{figure}

\begin{figure}[htbp]
\centerline{\psfig{figure=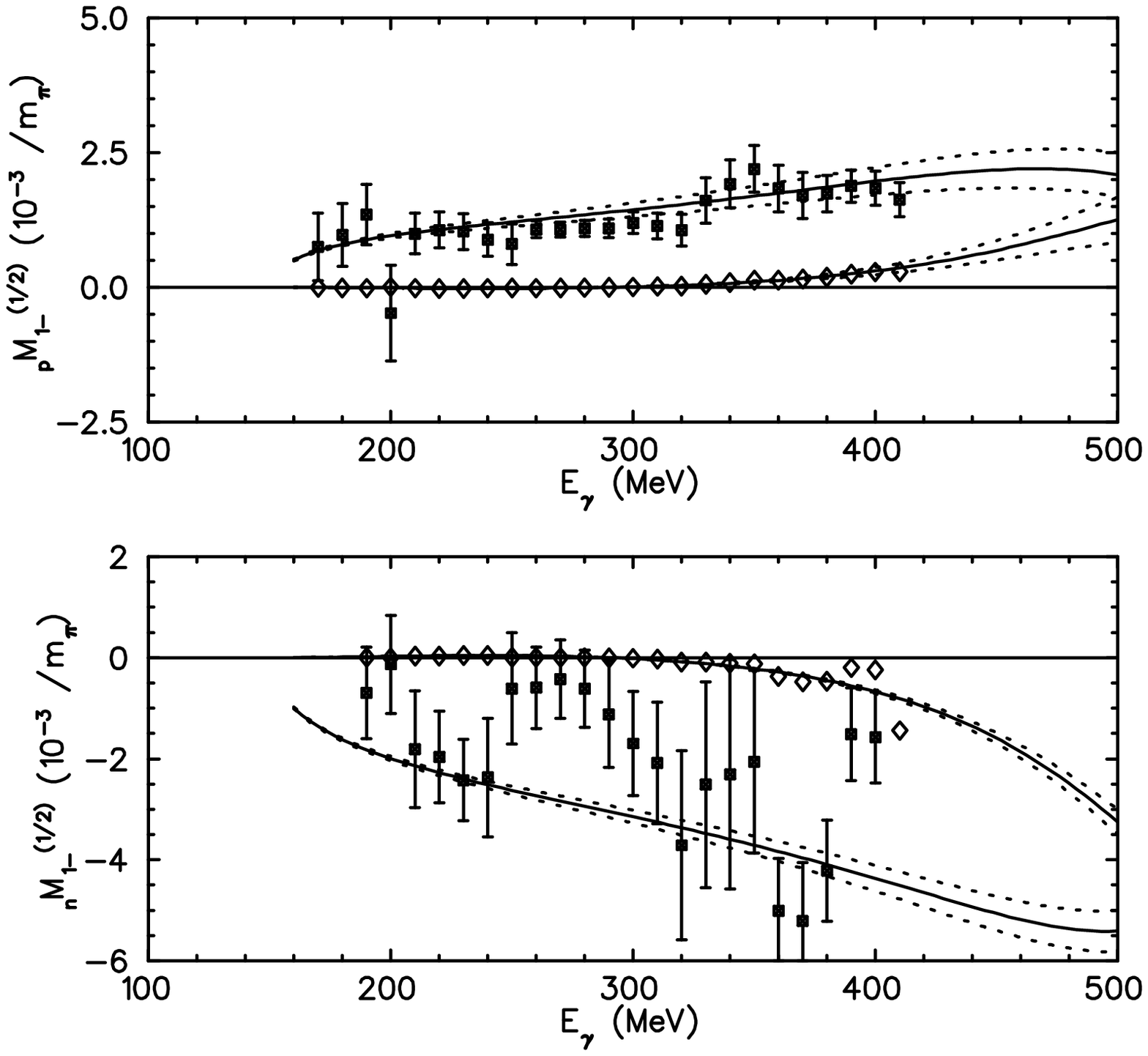,width=14cm}}
    \caption{Decomposition of the isospin $\frac{1}{2}$ channel of the
     amplitude $M_{1-}$ into components for reactions off the proton 
     and the neutron respectively. Symbols as in Fig.~\protect\ref{fig:13}.}
  \label{fig:18}
\end{figure}

\begin{figure}[htbp]
\centerline{\psfig{figure=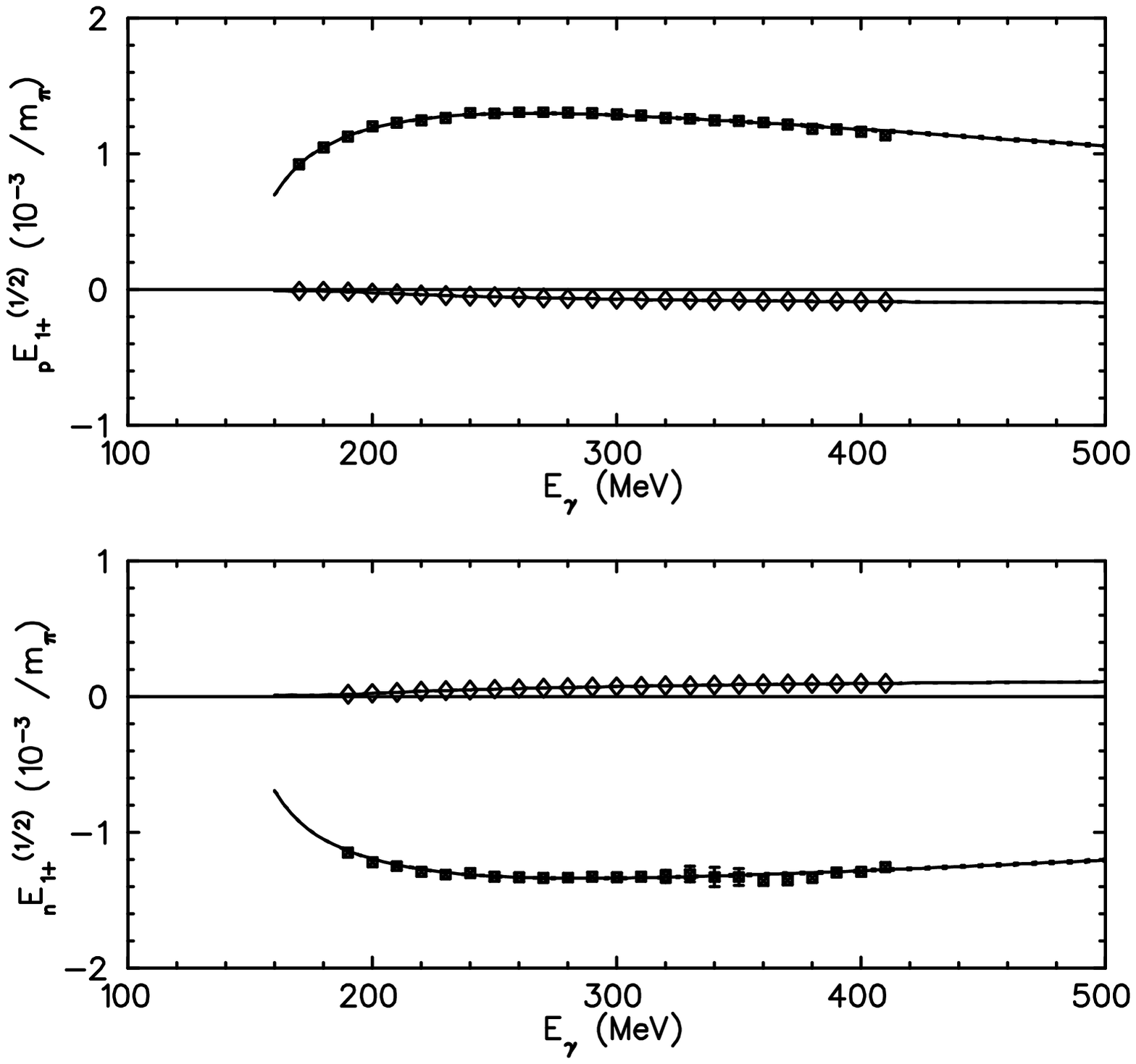,width=14cm}}
    \caption{Decomposition of the isospin $\frac{1}{2}$ channel of the
     amplitude $E_{1+}$ into components for reactions off the proton 
     and the neutron respectively. Symbols as in Fig.~\protect\ref{fig:13}.}
  \label{fig:19}
\end{figure}

\begin{figure}[htbp]
\centerline{\psfig{figure=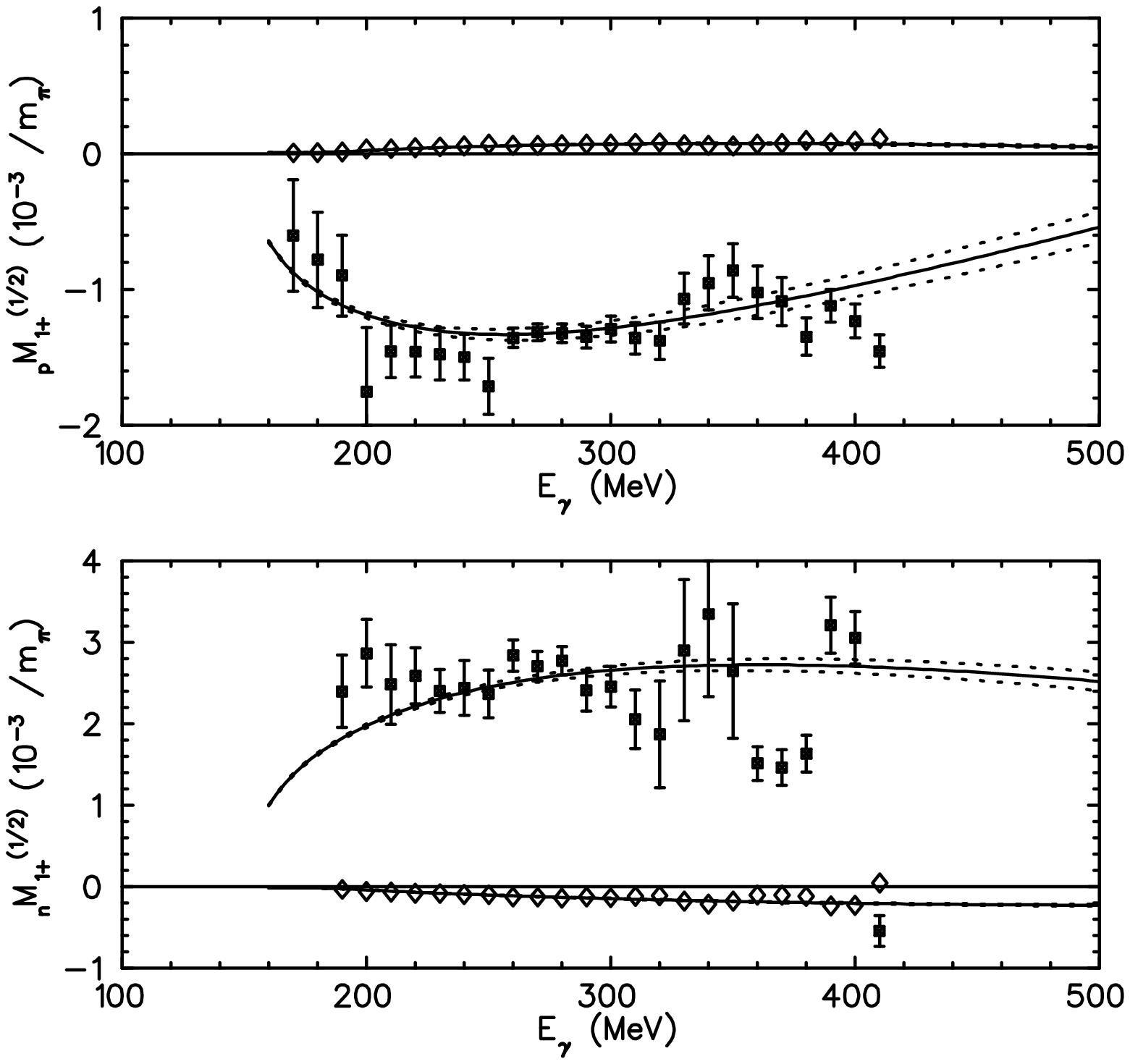,width=14cm}}
    \caption{Decomposition of the isospin $\frac{1}{2}$ channel of the
     amplitude $M_{1+}$ into components for reactions off the proton 
     and the neutron respectively. Symbols as in Fig.~\protect\ref{fig:13}.}
  \label{fig:20}
\end{figure}

In Figs.~\ref{fig:13}--\ref{fig:16} we show the 
isospin components of the $S$- and $P$-wave multipoles with 
$I = 0,\frac{1}{2},\frac{3}{2}$. The solid lines show the results for the real 
and imaginary parts of the ``global fit'' obtained with the complete 
data set. The error bands (dashed lines) result from the errors in 
the parameters as determined from the covariance matrix after normalizing
$\chi^{2}$ to 1. Because of the relatively small number of parameters, 
the errors of the global fit are very small except 
for the components of $E_{0+}$ and $M_{1-}$ with $I=0$, which reflects 
the uncertainty of the neutron data. In order to visualize the fluctuations of 
the data, we also analysed them in bins of 20 MeV with the same 
parametrization (``local fit''). Except for $\pi^{0}$ production on the 
proton, the data basis for energies $E_{\gamma} < 210$ MeV is quite 
unsatisfactory. As a result we find the largest error bars for all amplitudes 
in this region. Figs.~\ref{fig:17}--\ref{fig:20} show the decomposition of 
the isospin $\frac{1}{2}$ amplitudes into proton and neutron part according to 
Eqs.~(\ref{iso_iso_pn}). In this case, the corresponding 
local fit was obtained by first fitting the $_{p}M^{(\frac{1}{2})}$ and 
$M^{(\frac{3}{2})}$ components to the proton data, and then 
determining the $_{n}M^{(\frac{1}{2})}$
amplitudes by a fit to the neutron data. In the case of the 
proton, the local fit is in general agreement with the result 
of the global fit, while we find strong fluctuations and large error bars 
for the neutron amplitudes. It is obvious that more and better neutron 
data are needed to get full information on the 
amplitudes $E_{0+}$ and $M_{1-}$ in the isospin $\frac{1}{2}$ channel, which 
are of special interest for studies of the associated resonances $S_{11}$ 
and $P_{11}$. 

Given the results for the $S$ and $P$ waves from our fit, we now turn to 
a detailed study of the threshold amplitudes for both 
charged and neutral pions and to the electromagnetic excitation 
of the $\Delta(1232)$.

\subsection{Low energy amplitudes}
As pointed out before, our data basis does not include 
the threshold region. The main reason for this is that we have 
to expect isospin symmetry breaking in that region due to the pion 
mass splitting, while our dispersion calculation is based on the
isospin symmetry of the pion-nucleon phase shifts. Therefore, the 
threshold $S$ wave amplitudes given below are a prediction, in the 
sense that the cross sections above $E_{\gamma} = 160$ MeV determine 
the threshold values by continuation of the analytic 
functions constructed in dispersion theory \cite{Han97}. Table 
\ref{tab:2} compares our threshold values for the charged pion 
channels to the ``classical'' low energy theorem \cite{deB70} (LET), chiral
perturbation theory \cite{Ber96a,Ber96b} (ChPT) and experiment. Note 
that ChPT contains lowest order loop corrections, while LET is based 
on tree graphs only. The agreement between our calculation
and the experiment is excellent. 

\begin{figure}[htbp]
\centerline{\psfig{figure=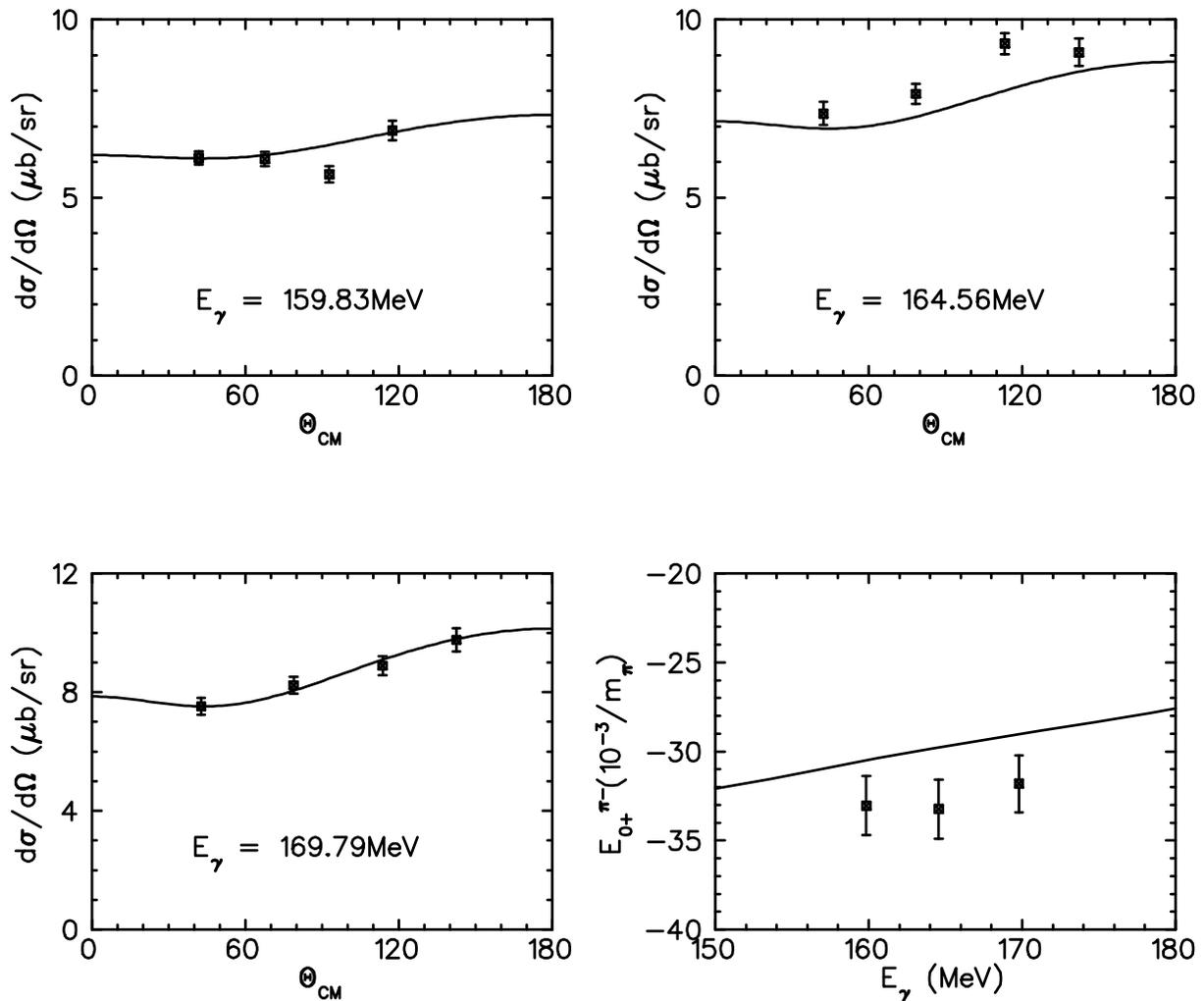,width=16cm}}
    \caption{Differential cross sections for $\pi^{-}$ production 
            at low energies. The results of our global fit compared to 
            the TRIUMF data \protect\cite{Liu94} for 
            $\gamma n\to\pi^{-}p$. The data points for $E_{0+}$ are also 
            from Ref.~\protect\cite{Liu94}.}
  \label{fig:pin_low}
\end{figure}

It should be noted that a  larger value for $E_{0+}(\pi^-p)$, namely 
$-34.7 \pm 1.0$ (here and in the following in units of 
$10^{-3}$/$m_{\pi}$), was reported \cite{Liu94}. However, this 
value derived from angular distributions was recently withdrawn \cite{Kov97}. 
Incidentally, that (wrong) value would have resulted in a difference 
of $a_1 - a_3 = 0.275/m_{\pi}$ for the pion-nucleon scattering lengths, while 
the (correct) value of Tab.~\ref{tab:2} gives a smaller 
value of $(0.253\pm 0.003)/m_{\pi}$. However, the larger value would be in much 
better agreement with the analysis of pion-nucleon scattering by the 
Karlsruhe group \cite{Hoe83} and recent results from pionic 
atoms which lead to a value of $(0.276\pm 0.013)/m_{\pi}$ \cite{Sig95}. 
Since this cross relation between hadronic
and electromagnetic physics is only based on the well-known Panofsky ratio, 
$\sigma(\pi^-p\rightarrow\pi^0n)$ / $\sigma(\pi^-p\rightarrow\gamma n)$, even 
the small deviation in $a_1-a_3$ poses a problem. Fig.~\ref{fig:pin_low} 
compares our results to the angular 
distributions near threshold measured at TRIUMF \cite{Kov97}. With the 
exception of the data at 165 MeV we find excellent agreement.

\begin{figure}[htbp]
\centerline{\psfig{figure=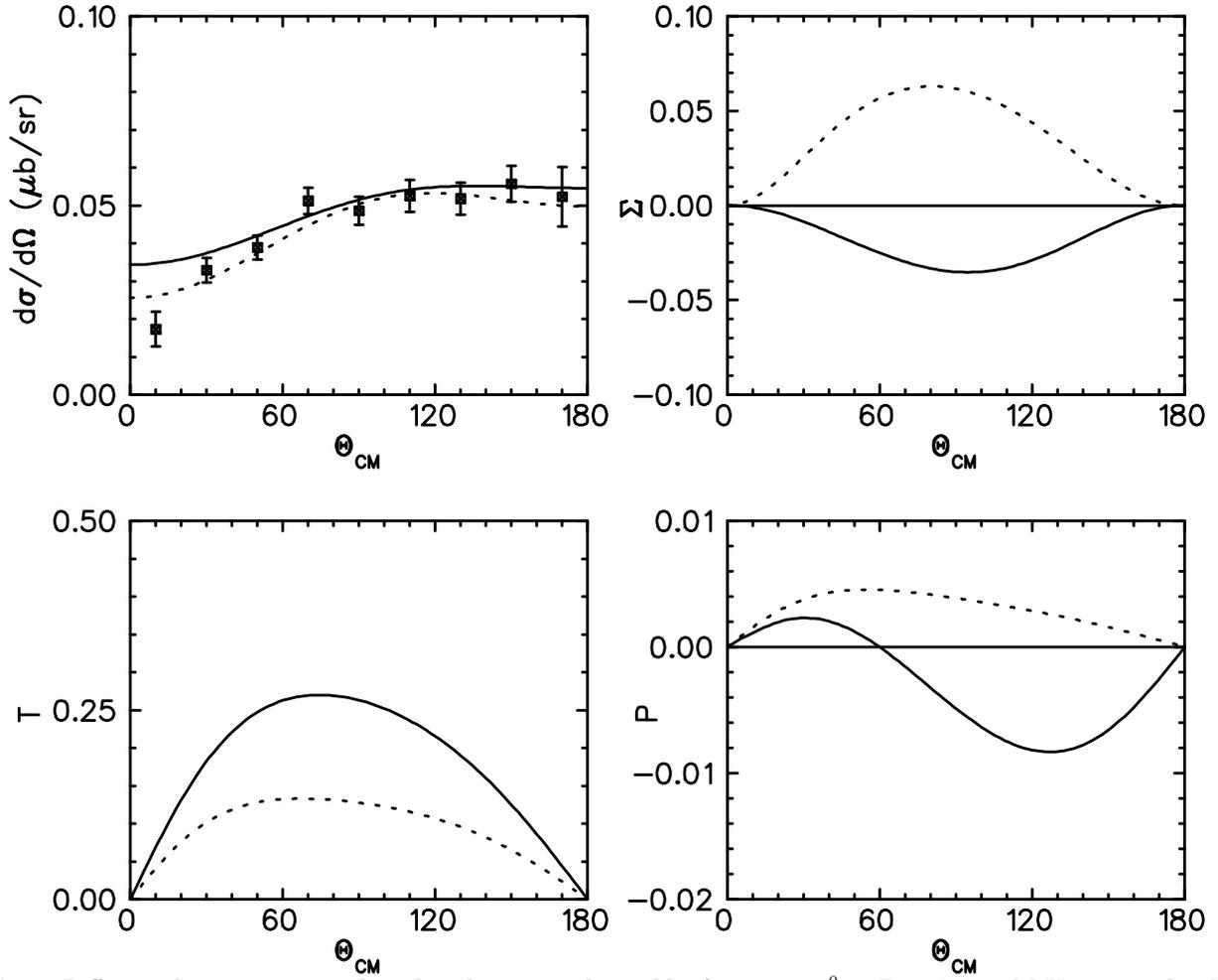,width=16cm}}
    \caption{Differential cross section and single polarization
      observables for $\gamma p\to p\pi^{0}$ at $E_{\gamma}=$ 151.59
      MeV compared to Mainz data \protect\cite{Fuc96}. The dotted line
      is obtained by reducing the contribution of the $M_{1-}$
      amplitude by 20 \% (see text).}
  \label{fig:low_pol}
\end{figure}

\begin{figure}[htbp]
\centerline{\psfig{figure=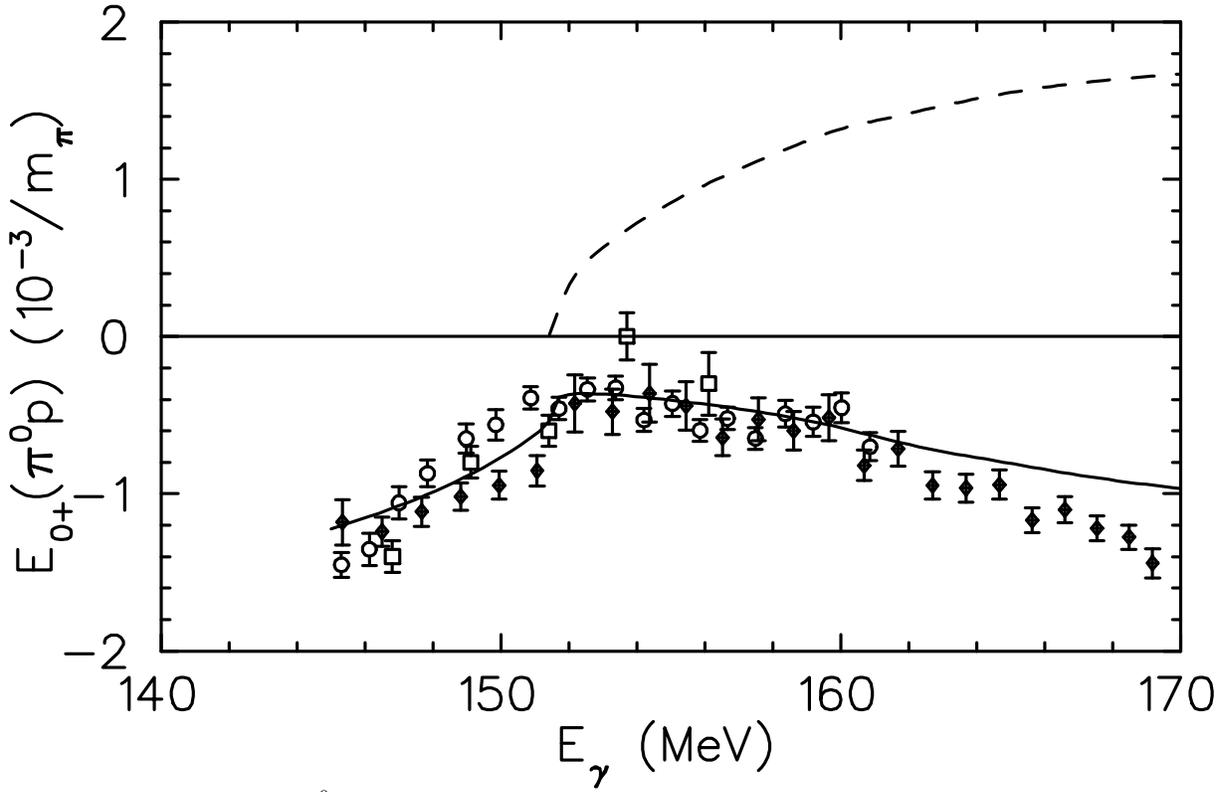,width=16cm}}
    \caption{The amplitude $E_{0+}(p\pi^{0})$ in comparison 
             with the recent analyses of data from Mainz 
             \protect\cite{Fuc96,Ber97}, 
             open circles, and Saskatoon \protect\cite{Ber96}, diamonds. The 
             open squares are the result of an analysis of an earlier 
             experiment at Mainz \protect\cite{Bec90,Dre92}. The 
             dashed line shows our prediction for Im$E_{0+}$, 
             the solid line for Re$E_{0+}$ respectively. The 
             unitarity parameter $\beta$  
             discussed in \protect\cite{Ber97} describes the 
             rise of Im$E_{0+}$ at $\pi^{+}$ threshold via 
             Im$E_{0+}=i\beta q_{+}$. In our calculation 
             $\beta =  4.3$, to be compared 
             with the values $3.76\pm 0.11$ corresponding to the unitary fit of 
             Ref.~\protect\cite{Ber97} and the ChPT value 2.78 
             \protect\cite{Ber96c}, all in units of $10^{-3}/m_{\pi^{+}}^{2}$.}
  \label{fig:e0p_low}
\end{figure}

Neutral pion photoproduction near threshold has been a topic of 
many experimental and theoretical investigations in the 90's. The 
classical LET predicted a power series expansion of the amplitude 
in $\mu = m_{\pi}/m_N \approx 1/7$, leading to a threshold value of 
about $-2.4$ plus presumably small corrections in $\mu^3$. The
experimental findings at Saclay \cite{Maz86}, Mainz \cite{Bec90} and 
now at Saskatoon \cite{Ber96} are a considerably different
amplitude of about half that value. The classical LET was based on 
Lorentz, gauge and chiral invariance \cite{deB70}. It further assumed that 
the amplitudes are analytical functions, though $\ln\mu$ singularities 
due to pion loops had been occasionally discussed in the literature. The first 
consistent calculation of pion photoproduction to one-loop order 
in ChPT clearly showed the breakdown of analyticity. The result is 
the expansion \cite{Ber91}
\begin{equation}
E_{0+}(p\pi^0) = \frac{eg_{\pi N}}{8\pi m_{\pi}}\left(\mu - 
                    \mu^{2}\frac{3 + \kappa_{p}}{2} 
                  - \mu^{2}\frac{m_{N}^{2}}{16f_{\pi}^{2}} + [\mu^3]\right),
\end{equation}
with the first two terms on the {\it rhs} given by the old LET, and the 
third term a correction due to pion loops. With the numerical 
values for the anomalous moment $\kappa_p$, the nucleon mass $m_N$, and 
the pion decay constant $f_\pi$ = 93 MeV, this correction even changes 
the sign of the threshold amplitude. Various predictions
are compared with the experiment in Tab.~\ref{tab:2}. The value 
given by ChPT was recently
evaluated in the heavy baryon formalism up to order $p^4$ \cite{Ber96b}. Our 
value from dispersion theory is obtained by breaking isospin symmetry 
explicitly by the physical masses of
the pions and by assuming that the imaginary parts of the amplitudes 
start at charged pion threshold and are given by the results of our 
global fit, i.e. by iteration of Eq.~(\ref{mul_int_fin}) under the stated 
assumptions. The agreement with experiment is astounding. Once more the 
data at the higher energies ``predict'' the threshold values with a good 
accuracy. This result is particularly surprising, because the 
real part of $E_{0+}$ obtains very large distributions from the imaginary 
parts of the higher multipoles via the dispersion integrals. Altogether 
these contributions nearly cancel the large contribution of the 
nucleon pole, which corresponds to the result of pseudoscalar 
coupling, leading to a total threshold value
\begin{equation} \nonumber
\mbox{Re}E_{0+}^{\mbox{\scriptsize thr}}(p\pi^0) 
 =  -7.63 + 4.15 - 0.41 + 2.32 + 0.29 + 0.07 = -1.22,
\end{equation}
\begin{equation} \nonumber
\mbox{Re}E_{0+}^{\mbox{\scriptsize thr}}(n\pi^{0})
 = -5.23 + 4.15 - 0.41 + 3.68 - 0.93 - 0.05 = 1.19,
\end{equation}
where the individual contributions on the $rhs$ are, in that 
order, the pole term, $M_{1+}, E_{1+}, E_{0+}, M_{1-}$ and higher 
multipoles. We add that the existing uncertainties for the neutron 
value of $M_{1-}$ give rise to large effects. Fig.~\ref{fig:low_pol} 
compares our
results to the measured angular distributions and predicts values for 
the three single polarization observables. The dotted curves are obtained 
by arbitrarily reducing $M_{1-}$ by 50 \%, which demonstrates the strong 
sensitivity of $\Sigma$ and P with regard to that multipole. Finally, 
Fig.~\ref{fig:e0p_low}  shows the energy dependence of $E_{0+}$ in the 
threshold region. The existence of a Wigner cusp in Re$E_{0+}(p\pi^0)$ at 
charged pion threshold is due to the infinite derivative of Im$E_{0+}$ at 
that threshold, and a reflection of the coupling to the much stronger 
$\pi^+$ channel. We also calculated the threshold values for the 
three $P$ waves and find agreement with the prediction of ChPT  and
experimental data within 5-10 \%. For a detailed discussion see 
Ref.~\cite{Han97}.

\subsection{The ratio $E_{1+}/M_{1+}$ in the $\Delta$ region}

The search for a deformation of the ``elementary'' particles 
is a longstanding issue. Such a deformation is evidence 
for a strong tensor force between the constituents,
originating in the case of the nucleon from the residual 
force of gluon exchange between the quarks. Depending on one's favourite 
model, such effects can be described by d-state
admixture in the quark wave function, tensor correlations 
between the pion cloud and the quark bag, or by exchange currents 
accompanying the exchange of mesons between the
quarks. Unfortunately, it would require a target with a spin of at 
least 3/2 (e.g. $\Delta$ matter) to observe a static deformation. The 
only realistic alternative is to measure the transition quadrupole 
moment between the nucleon and the $\Delta$, i.e. the amplitude 
$E_{1+}$. While the existence of such a transition cannot be regarded as
direct proof of a deformation, it is sensitive to model parameters 
responsible for possible deformations of the hadrons. 

The experimental quantity of interest is the ratio 
$R_{EM} = E_{1+}/M_{1+}$ in the region of the $\Delta$. The quark 
model with SU(6) symmetry predicts $R_{EM} = 0$. Depending on the size 
of the hyperfine interaction and the bag radius, broken SU(6)
leads to $-2 \% < R_{EM} < 0$ \cite{Kon80,Dre84}. Similar results 
can be obtained in the cloudy bag model \cite{Kae83,Ber88}, larger effects 
with Skyrmions \cite{Wir87} ($-5 \% < R_{EM} < -2 \%$). Exchange 
currents \cite{Buc96} yield values of about $-3.5 \%$. The first lattice
QCD results are ($+3 \pm 9$) \% \cite{Lei92}. The analysis of the data 
changes by time and method, with a PDG average 
of ($-1.5 \pm 0.4$) \% \cite{PDG96}.

\begin{figure}[htbp]
\centerline{\psfig{figure=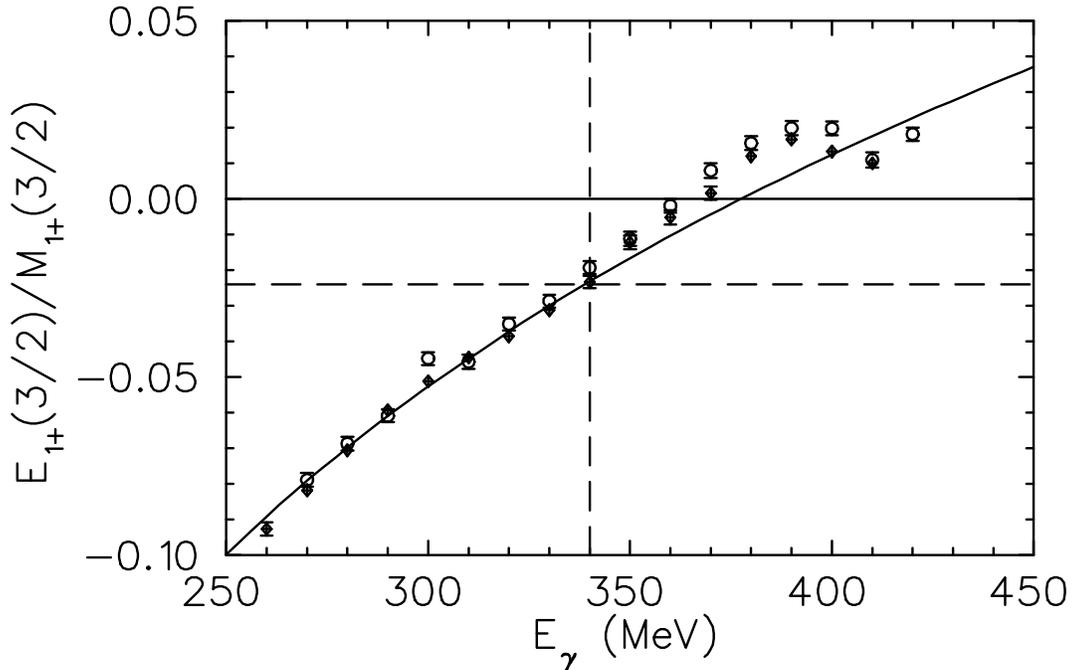,width=14cm}}
    \caption{The ratio $E_{1+}^{(\frac{3}{2})}/M_{1+}^{(\frac{3}{2})}$ as 
             a function of $E_{\gamma}$. Solid line: global fit, 
             diamonds: local fit, circles: analysis 
             of Krahn \protect\cite{Kra96}.}
  \label{fig:e_m}
\end{figure}

\begin{figure}[htbp]
\centerline{\psfig{figure=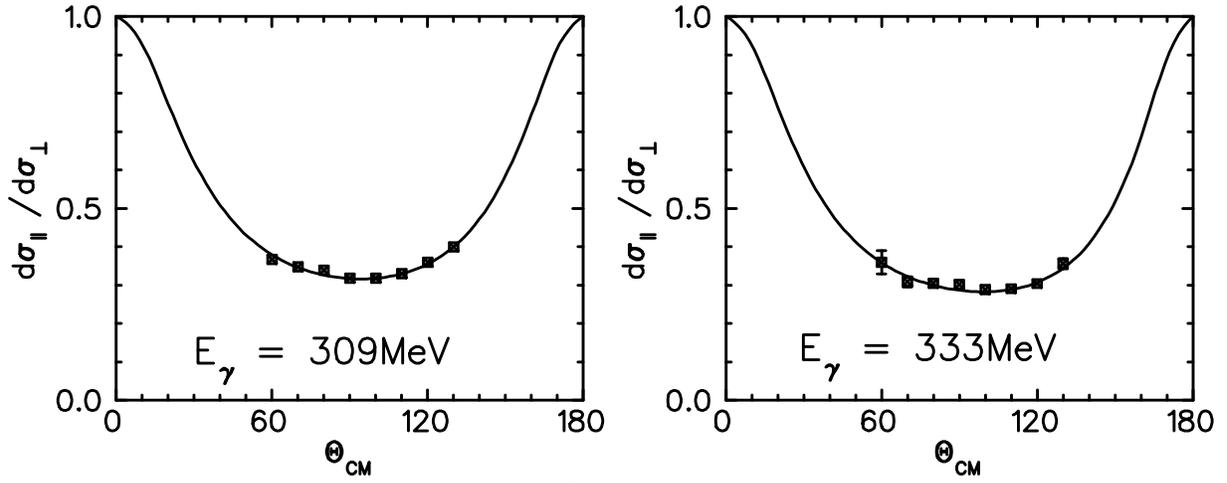,width=16cm}}
    \caption{The ratio $d\sigma_{\parallel}/d\sigma_{\perp}$ 
             for $\gamma p\to p\pi^{0}$ compared to the analysis of Sandorfi 
             et al.\protect\cite{San96}.}
  \label{fig:legs}
\end{figure}

The comparison of our dispersion analysis \cite{Han96} with the analysis 
of the Mainz data by Beck et al. \cite{Bec95}
shows nice agreement with the multipole $M_{1+}$ but 
systematic differences for the quadrupole transition $E_{1+}$ at energies
above the resonance. However, the ratio $R_{EM}$ at resonance agrees 
quite well in both analyses (see Fig.~\ref{fig:e_m}). Unfortunately, there 
appears to be a discrepancy of up to 15 \% between the differential cross 
sections measured at Mainz (upon which our analysis is largely based) and 
the LEGS data \cite{Bla97} at Brookhaven. As far as ratios of cross
sections are concerned, the data seem to be in general agreement. As an 
example, the ratio of the LEGS cross sections for photons with polarization 
parallel and orthogonal to the scattering plane \cite{San96}, is 
nicely described by our analysis (see Fig.~\ref{fig:legs}), though 
these data have not been included in our fit. In view of the importance of 
the quantity $R_{EM}$ for our understanding of the internal structure of 
the nucleon,the existing discrepancies between the data deserve further 
studies. Based on ``our'' data selection we obtain a value $R_{EM}$ =
($-2.5 \pm 0.1)$ \% for the global fit ($(-2.33 \pm 0.17)$ \% for 
the local fit), which is close to the value 
derived by Beck et al.\cite{Bec97}.

Another point of interest is the separation of resonance and background 
contributions. The problem in this quest arises, because the leading 
tree graphs, the Born terms and the $\Delta$ excitation, do not fulfill 
the requirements of unitarity (Fermi-Watson theorem). In physical terms, the 
pion produced by the Born graph feels the presence of
the resonance and is reabsorbed by the nucleon to form a $\Delta$, which 
altogether leads to the required phase relation of Eq.~(\ref{multi_wat}). 
Together with the
``pure'' $\Delta$, these diagrams constitute the ``dressed''
$\Delta$. A technique to achieve this result is the method of
Olsson \cite{Ols74},
\begin{equation}
M_{1+} = |M_{1+}^B|e^{i \delta_B} +|M_{1+}^{\mbox{\scriptsize res}}|e
^{i \delta_{\mbox{\scriptsize res}}}e^{i \Phi} =
|M_{1+}|e^{i \delta_{1+}},
\end{equation}
which combines the two contributions, unitarized individually, by a rotation 
of the resonance contribution in the complex plane. The other technique is the
K-matrix method \cite{Dav90},
\begin{equation} 
K(W) = A(W)/(M_{\mbox{\scriptsize res}}-W) + B(W),
\end{equation}
\begin{equation} 
T(W) = K(W) cos \delta (W) e^{i \delta (W)},
\end{equation}
where $B(W)$ is a slowly varying background, and 
$\delta (W = M_{\mbox{\scriptsize res}}) = \pi/2$.
It is obvious that different approaches will 
give different answers to the question which 
part of ratio $R_{EM}$ is due to the ``pure'' resonance.

Due to the analyticity of the functions 
involved, dispersion theory is particularly apt to
study the amplitudes in the complex energy plane. Therefore, we 
have analyzed the amplitude by comparing the ``speed'' \cite{Hoe92}
\begin{equation}
SP\left[M_{1+}(W)\right] = \left | \frac{dM_{1+}(W)}{dW} \right |
\end{equation}
with the speed of an ideal resonance due to a pole in the complex plane at 
$W_R = M_R-i \Gamma_R/2$. Such a resonance has the 
amplitude
\begin{equation}
M_{1+}^{\mbox{\scriptsize res}}(W) = 
\frac{re^{i \Phi}\Gamma_{R}}{M_R-W-i \Gamma_{R}/2}\hspace{3mm} ,
\end{equation}
with $re^{i \Phi}$ determining the complex residue at resonance, $W = W_R$.

\begin{figure}[htbp]
\centerline{\psfig{figure=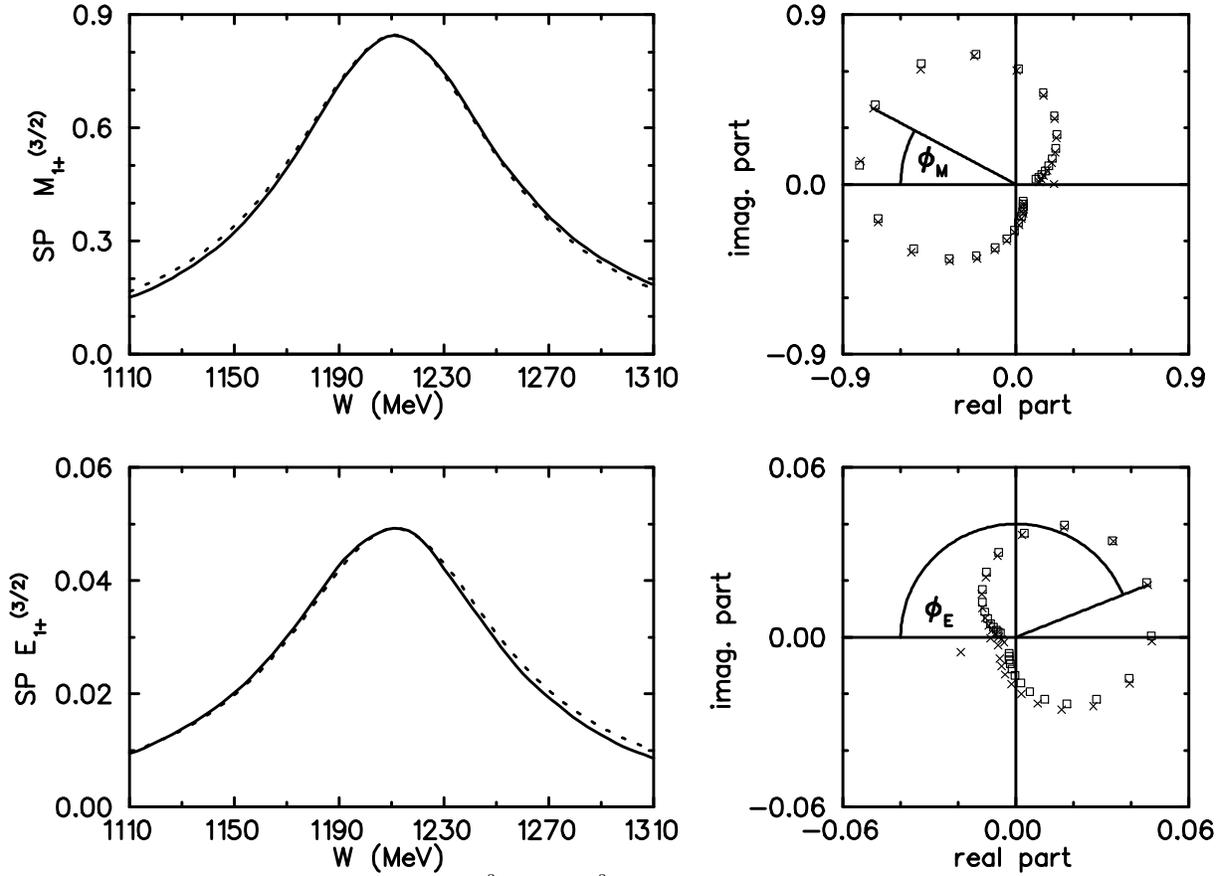,width=16cm}}
    \caption{Left: The speed of the multipoles $M_{1+}^{(\frac{3}{2})}$ and 
            $E_{1+}^{(\frac{3}{2})}$ (solid line) compared to the case of 
            an ideal resonance pole (dotted line). Right: The speed 
            vectors for these multipoles in the complex plane 
            (crosses) compared to the case of 
            an ideal resonance pole (squares).}
  \label{fig:sp}
\end{figure}

\begin{figure}[htbp]
\centerline{\psfig{figure=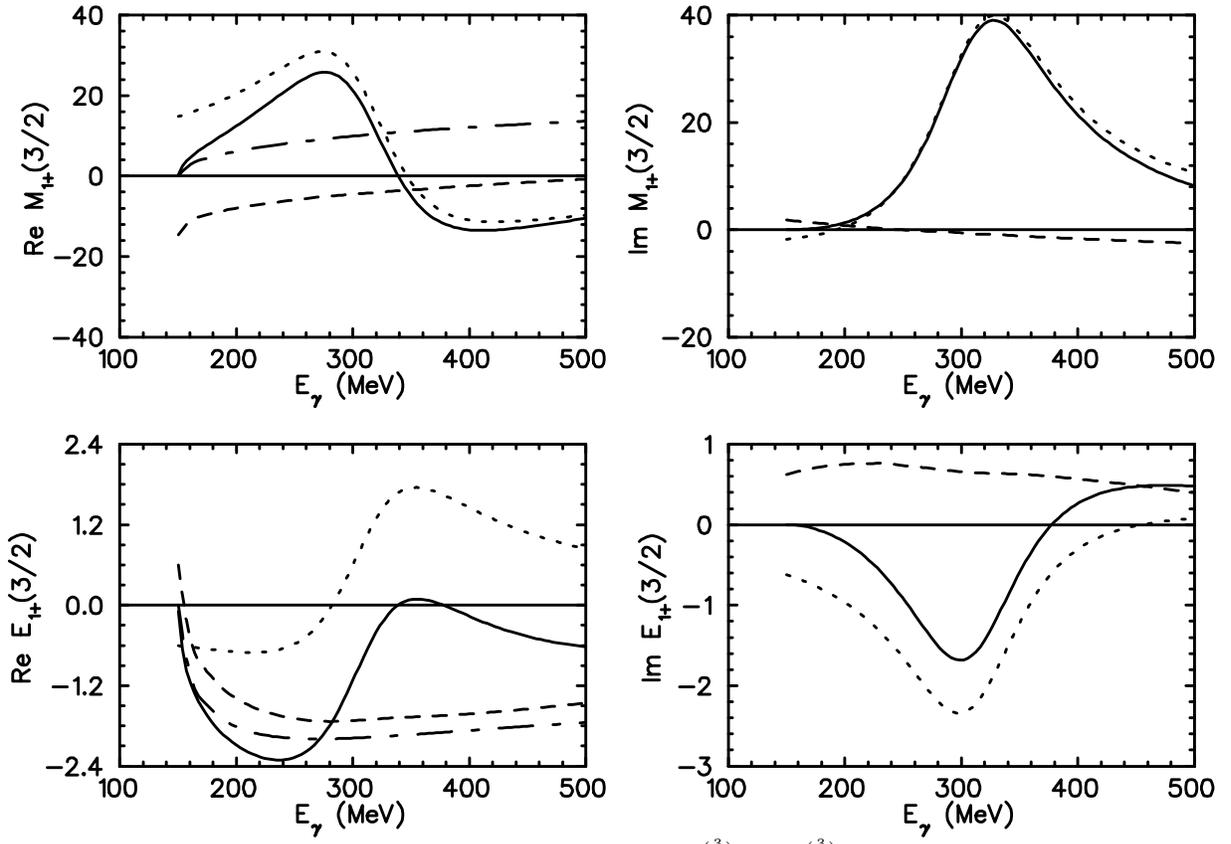,width=16cm}}
    \caption{Separation of resonance and background contributions 
             for  $M_{1+}^{(\frac{3}{2})}$ and $E_{1+}^{(\frac{3}{2})}$ into 
             full amplitude (solid lines), resonance pole (dotted lines), 
             background (dashed lines) and Born terms only 
             (dashed-dotted lines).}
  \label{fig:uni}
\end{figure}

Due to the differentiation involved, the effect of a slowly varying 
background is much reduced, and the ``experimental'' speed agrees 
nicely with the ``ideal'' speed, as can be seen in Fig.~\ref{fig:sp} 
(left part). The same is true for the speed vector, the complex number
$dM_{1+}/dW$, which has been plotted on the $rhs$ of Fig.~\ref{fig:sp}. The 
experimental values given by the crosses in this figure agree quite 
well with the squares of the ideal resonance, except for the smallest 
energies close to the origin. We note that as function of energy W the 
values run counter-clockwise, starting near the origin at
threshold (the ``stem'' of the apple-shaped structure), passing through 
resonance at maximum speed (the ``blossom'' of the apple), and again 
approaching the origin for $W \rightarrow \infty$. The fact that the ``apples'' 
for $M_{1+}$ (top) and $E_{1+}$ (bottom) are essentially oriented in 
opposite directions is related to the negative value of $R_{EM}$. The 
phases of the complex amplitudes at resonance are defined by
the angles $\Phi_M$ and $\Phi_E$ respectively 
(see Fig.~\ref{fig:sp}). Table \ref{tab:res} shows 
our results for the pole position and the complex residue at the pole 
for both the magnetic dipole and the electric quadrupole transitions. The 
complex ratio of the residues is given by
\begin{equation}
R_{EM}^{\Delta} = \frac{re^{i \Phi}(E_{1+})}{re^{i \Phi}(M_{1+})} 
= -0.035-0.046i.
\end{equation}
The values for the pole position are in perfect agreement with the 
analysis of pion-nucleon scattering. Essentially based on 
analyses of H\"ohler et al., the PDG \cite{PDG96} lists 
$M_R = (1210 \pm 1)$ MeV and $\Gamma_R = (100 \pm 2)$ MeV. The complex 
ratio $R_{EM}^\Delta$
at the $\Delta$ pole, $W = M_R-i \Gamma_{R}/2$, is rather stable in 
comparison with the value $R_{EM}$ on the real axis, at 
$W = M_{\mbox{\scriptsize res}} \approx 1232$ MeV. As an example, Sato et 
al.~\cite{Sat96} have studied two versions of a dynamical model leading to a
difference in $R_{EM}^{\Delta}$ of 30 \% while $R_{EM}$ differs by as 
much as 300 \%! It is also worth noting that studies of hyperon 
resonance radiative decay in the framework of
ChPT have predicted values for $R_{EM}^\Delta$ compatible with our
results within the error bars. Finally, Fig.~\ref{fig:uni} 
shows the resonance-background separation
of the real and imaginary parts for the two $\Delta$ 
multipoles. Indeed, the speed-plot technique leads to a slowly varying 
background except for the threshold region where the background has
to cancel the ``ideal'' resonance shape to give the (vanishing) physical 
amplitude. While the experimental amplitude $E_{1+}$, particularly its 
real part, is quite different from a typical resonance shape, the 
amplitude after background subtraction is not too far from an
ideal resonance. It is in fact rather astounding that $M_R$ 
can be predicted to four digits and $\Gamma_R$ to two digits from 
the $E_{1+}$ amplitude, which is about half resonance
and half background (see Fig.~\ref{fig:uni}).

Instead of the multipoles the electromagnetic resonance excitation 
is very often discussed in terms of helicity amplitudes. In 
Tab.~\ref{tab:hel} we give the helicity matrix elements $A_{\frac{1}{2}}$ 
and $A_{\frac{3}{2}}$ for the transition $\gamma N\to\Delta$ which are derived 
from our results for the multipoles via the standard definitions. In that 
table also the ratio $R_{EM}$ is given which is equivalent to the 
result of the application of the K-matrix method.

\subsection{Total cross sections}
\begin{figure}[htbp]
\centerline{\psfig{figure=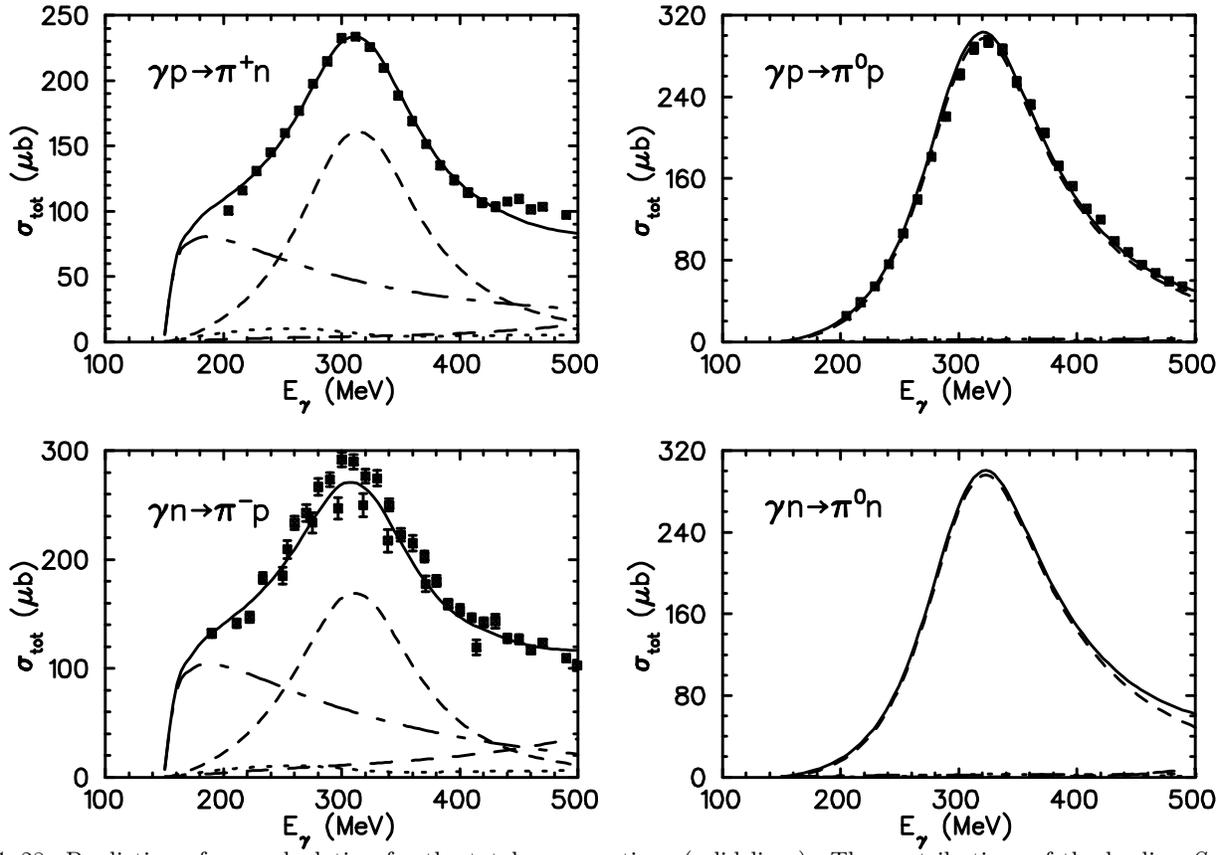,width=16cm}}
    \caption{Prediction of our calculation for the total cross sections 
             (solid lines). The contributions of the leading 
              $S$ and $P$ wave multipoles are also shown: $M_{1+}$ 
              (dashed lines), $E_{0+}$ (dashed-dotted lines), 
              $M_{1-}$ (long dashed lines), $E_{1+}$ (dotted lines). 
              The data are from Mainz \protect\cite{Hae96,A2-95} for $\pi^{+}$ 
              and $\pi^{0}$, and from  Tokyo \protect\cite{Fuj72} 
              and Frascati \protect\cite{Chi75} for $\pi^{-}$.}
  \label{fig:sig_tot}
\end{figure}

As a result of our fit we also obtain predictions for the total cross sections. 
These are shown in Fig.~\ref{fig:sig_tot} for the four physical reactions, 
together with the contributions of the leading $S$- and $P$-wave 
multipoles. In terms of the multipoles, the total cross sections are 
\begin{equation}
  \label{sig_sum}
  \sigma_{{\mbox tot}} = 2\pi\frac{\mid\vec k\mid}{\mid\vec q\mid}
  \sum_{l=0}^{\infty}(l+1)^{2}\left[(l+2)\left(\mid E_{l+}\mid^{2}
  +\mid M_{l+1,-}\mid^{2}\right)+l\left(\mid M_{l+}\mid^{2}
  +\mid E_{l+1,-}\mid^{2}\right)\right].
\end{equation}
with the isospin combinations of the multipoles according to Eqs.
(\ref{iso_phys_1}--\ref{iso_phys_4}).

\begin{figure}[htbp]
\centerline{\psfig{figure=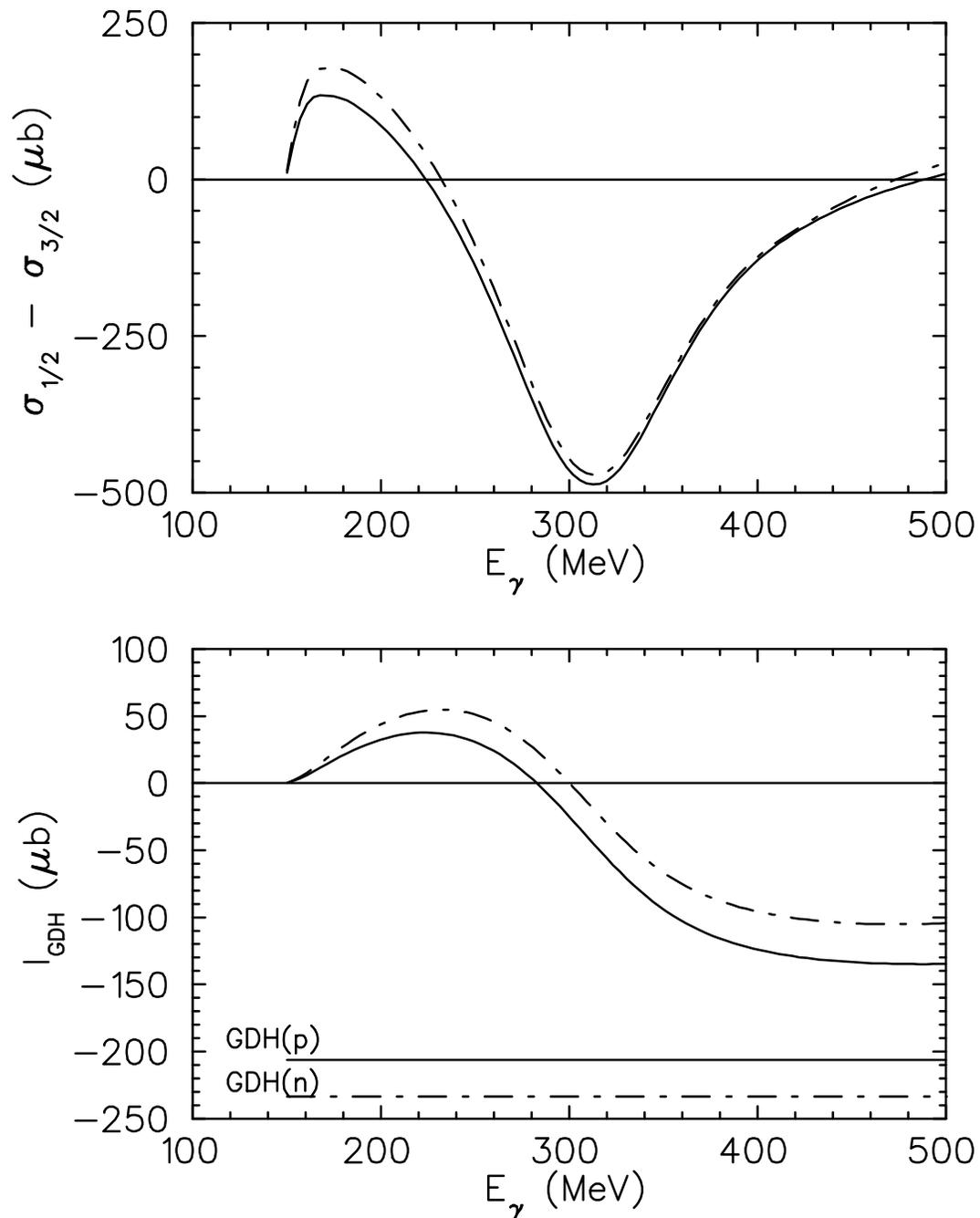,width=14cm}}
    \caption{Our prediction for the difference 
             $\sigma_{\frac{1}{2}}-\sigma_{\frac{3}{2}}$ on the proton 
             (solid line) and the neutron (dashed-dotted line) and the 
             corresponding contributions to the GDH integral, which 
             are evaluated as functions of the upper limit of integration. 
             We also indicate the predictions of the sum rule.}
  \label{fig:GDH}
\end{figure}

A topic of much current interest is the spin structure of the total absorption 
cross section, which will be measured with circularly polarized photons 
and nucleon target polarization in the direction of the photon 
momentum \cite{MAM93,ELS93}. Depending on their relative 
direction, the spins of photon and 
nucleon add to spin 3/2 or subtract to spin 1/2, which defines the 
helicity structure of the total cross sections, $\sigma_{\frac{3}{2}}$ and 
$\sigma_{\frac{1}{2}}$, with $\sigma_{{\mbox tot}}=(\sigma_{\frac{3}{2}}
+\sigma_{\frac{1}{2}})/2$. The difference of these two qualities enters into 
the Gerasimov-Drell-Hearn (GDH) sum rule,
\begin{equation}
  \label{DHG_sum}
  -\frac{\kappa^{2}}{4}=\frac{m_{N}^{2}}{8\pi^{2}\alpha}
  \int\limits_{\nu_{thr}}^{\infty}\frac{d\nu}{\nu}
  \left(\sigma_{\frac{1}{2}}(\nu)-\sigma_{\frac{3}{2}}(\nu)\right),
\end{equation}
which relates the helicity structure of the total absorption cross section of 
the nucleon with its ground state properties. Our prediction for the 
contributions of single pion production to 
$\sigma_{\frac{3}{2}}-\sigma_{\frac{1}{2}}$ and the GDH integral as function 
of its upper limit $E_{\gamma}$ is shown in Fig.~\ref{fig:GDH}.

\section{Summary and outlook}

Dispersion relations at fixed $t$ provide a useful framework 
to analyze pion photoproduction between threshold and an 
excitation energy of about 500 MeV. After multipole projection we obtain a 
system of coupled integral equations for the multipole 
amplitudes is whose solutions depend on 10 free parameters, which 
are fitted to selected experimental data in the range of 
160 MeV $\leq E_{\gamma}\leq$ 420 MeV. On the basis of these amplitudes, we 
obtain predictions for the threshold behaviour of pion 
photoproduction, including the 
cusp effect for neutral pion production. The results 
for $S$ and $P$ waves agree 
within about 5 - 10 \% with the predictions of ChPT and a direct analysis
of the threshold data.

In the region of the $\Delta$ (1232), we find a ratio of the electric 
quadrupole to the magnetic dipole amplitude of $R_{EM} = (-2.5 \pm 0.1) \%$ 
at $W = M_{\mbox{\scriptsize res}} = 1232$ MeV. The $\Delta$ resonance pole 
in the complex plane can be determined from both 
amplitudes, in excellent agreement with the results of pion-nucleon 
scattering. The ratio of the complex residues at the pole is $R_{
EM}^{\Delta} = (-3.5 - 4.6i) \%$. While $R_{EM}$ is very model 
dependent, the value at the resonance pole is relatively stable.

Since our ``global fit'' contains only relatively few parameters, it 
cannot follow all the fluctuations of the data in the energy and angular 
distributions. Therefore, the overall $\chi^2$ is only 2.8 per data 
point. We are convinced that many of the fluctuations of
the data are due to systematical errors and, therefore, should not be 
reproduced by a theoretical analysis. In order to obtain an estimate 
of such systematical errors, we also show the result of a  ``local fit'' 
by fitting the free parameters to, e.g., the data in 20 MeV bins 
individually. 

As further tests of the remaining model dependence of our 
procedure, we are presently 
studying some modifications of the parametrization. First, the high 
$t$-channel contributions to the dispersion integrals will be replaced 
by a Regge model. Second, all isospin components of the $S$, $P$ and $D$ 
waves will be treated on the same footing by
allowing for the addition of homogeneous solutions to all of them. 
The explicit fit of  
all $D$ wave contributions will give the possibility to study the influence 
of the higher partial waves.

In conclusion we have obtained a good representation of the data 
in the range up to about 500 MeV. The biggest experimental uncertainties are 
shown to be due to the neutron data, in particular in the case of 
the $E_{0+}$ and $M_{1-}$ amplitudes. New experiments to determine the
small amplitudes by polarization observables are urgently needed, 
particularly in the case of the neutron and the Roper multipoles.

\appendix

\section{Matrices connecting the amplitudes}
\label{matrices}
The matrices connecting the various amplitudes defined in the text are:
\begin{equation}
  \label{Mat_H}
  {\mbox{\bf H}}  =  diag\left(\begin{array}{c}
    \frac{\sqrt{(E_{f}+m_{N})(E_{i}+m_{N})}(W-m_{N})}{8\pi W}\\
    \frac{\sqrt{(E_{i}+m_{N})}}{\sqrt{E_{f}+m_{N}}}
          \frac{(W-m_{N})\mid\vec k\mid}{8\pi W}\\
    \frac{\sqrt{(E_{f}+m_{N})(E_{i}+m_{N})}(W-m_{N})\mid\vec k\mid}{8\pi W}\\
    \frac{\sqrt{E_{i}+m_{N}}}{\sqrt{E_{f}+m_{N}}}
          \frac{(W-m_{N})\mid\vec k\mid^{2}}{8\pi W}
  \end{array}\right),\nonumber
\end{equation}
\begin{equation}
  \label{Mat_C}
  {\mbox{\bf C}} = \left(
    \begin{array}{cccc}
 1 & 0 & -\frac{t-m_{\pi}^{2}}{2(W-m_{N})} 
   & W-m_{N}+\frac{t-m_{\pi}^{2}}{2(W-m_{N})} \\
-1 & 0 & -\frac{t-m_{\pi}^{2}}{2(W+m_{N})} 
   & W+m_{N}+\frac{t-m_{\pi}^{2}}{2(W+m_{N})} \\
 0 & W-m_{N} & 1 & -1 \\ 0 & -(W+m_{N}) & 1 & -1
    \end{array}\right),
\end{equation}
\begin{equation}
  \label{Mat_C_inv}
  {\mbox{\bf C}}^{-1} = \frac{1}{4W^{2}} \left(
    \begin{array}{cccc}
 (C^{-1})_{11} & (C^{-1})_{12} & (C^{-1})_{13} & (C^{-1})_{14} \\
 0 & 0 & 2W & -2W \\
  2W & 2W & (C^{-1})_{33} & (C^{-1})_{34} \\
  2W & 2W & W\frac{t-m_{\pi}^{2}}{W-m_{N}} & W\frac{t-m_{\pi}^{2}}{W+m_{N}}
    \end{array}\right),
\end{equation}
with
\begin{eqnarray*}
  (C^{-1})_{11} & = & 2W(W+m_{N}), \\
  (C^{-1})_{12} & = & -2W(W-m_{N}), \\
  (C^{-1})_{13} & = & 2m_{N}W\frac{t-m_{\pi}^{2}}{W-m_{N}}, \\
  (C^{-1})_{14} & = & 2m_{N}W\frac{t-m_{\pi}^{2}}{W+m_{N}}, \\
  (C^{-1})_{33} & = & 
      2W\left(W+m_{N}+\frac{t-m_{\pi}^{2}}{2(W-m_{N})}\right), \\
  (C^{-1})_{34} & = & 
      2W\left(W-m_{N}+\frac{t-m_{\pi}^{2}}{2(W+m_{N})}\right). \\
\end{eqnarray*}

\begin{equation}
  \label{Mat_Proj}
  {\mbox{\bf G}}_{l}(x) = \left(
    \begin{array}{cccc}
      P'_{l+1}(x) & P'_{l-1}(x) & lP'_{l+1}(x) & (l+1)P'_{l-1}(x) \\
      0 & 0 & (l+1)P'_{l}(x) & lP'_{l}(x) \\
      P''_{l+1}(x) & P''_{l-1}(x) & -P''_{l+1}(x) & P''_{l-1}(x) \\
     -P''_{l}(x) & -P''_{l}(x) & P''_{l}(x) & -P''_{l}(x) 
    \end{array}\right),
\end{equation}
\begin{small}
\begin{equation}
  {\mbox{\bf D}}_{l}(x) = \left(
  \begin{array}{cccc}
   \frac{1}{2(l+1)}\{P_{l} & -P_{l+1} & \frac{l}{2l+1}(P_{l-1}-P_{l+1})
   & \frac{l+1}{2l+3}(P_{l}-P_{l+2})\} \\
   \frac{1}{2l}\{P_{l} & -P_{l-1} & \frac{l+1}{2l+1}(P_{l+1}-P_{l-1})
   & \frac{l}{2l-1}(P_{l}-P_{l-2})\} \\
   \frac{1}{2(l+1)}\{P_{l} & -P_{l+1} & \frac{1}{2l+1}(P_{l+1}-P_{l-1})
   & 0 \} \\
   \frac{1}{2l}\{-P_{l} & P_{l-1} & \frac{1}{2l+1}(P_{l-1}-P_{l+1})
   & 0 \}
  \end{array}\right),
\end{equation}
\end{small}
with the Legendre polynomials $P_{l}$ as functions of $x$.

The entries of the matrix $\mbox{\bf{E}}$ are:
\begin{eqnarray}
  \label{E_Mat}
  E_{11}^{I} & = & \frac{1}{W'-W}+\frac{\varepsilon^{I}}{W'^{2}-u}\left(W'+W+
           \frac{t-m_{\pi}^2}{W-m_{N}}\right), \nonumber \\
  E_{12}^{I} & = & -\frac{1}{W'+W}+\frac{\varepsilon^{I}}{W'^{2}-u}\left(W-W'+
           \frac{t-m_{\pi}^2}{W-m_{N}}\right), \nonumber \\
  E_{13}^{I} & = & -\frac{1-\varepsilon^{I}}{2}
                \frac{t-m_{\pi}^{2}}{(W'-m_{N})(W-m_{N})}, \nonumber \\
  E_{14}^{I} & = & -\frac{1-\varepsilon^{I}}{2}
                \frac{t-m_{\pi}^{2}}{(W'+m_{N})(W-m_{N})}, \nonumber \\
  E_{21}^{I} & = & -\frac{1}{W'+W}+\frac{\varepsilon^{I}}{W'^{2}-u}\left(W-W'+
           \frac{t-m_{\pi}^2}{W+m_{N}}\right), \nonumber \\ 
  E_{22}^{I} & = & \frac{1}{W'-W}-\frac{\varepsilon^{I}}{W'^{2}-u}\left(W'+W+
           \frac{t-m_{\pi}^2}{W+m_{N}}\right), \nonumber \\
  E_{23}^{I} & = & -\frac{1-\varepsilon^{I}}{2}
                \frac{t-m_{\pi}^{2}}{(W'-m_{N})(W+m_{N})}, \nonumber \\
  E_{24}^{I} & = & -\frac{1-\varepsilon^{I}}{2}
                \frac{t-m_{\pi}^{2}}{(W'+m_{N})(W+m_{N})}, \\
  E_{31}^{I} & = & -\frac{2\varepsilon^{I}}{W'^{2}-u}, \nonumber \\
  E_{32}^{I} & = & -\frac{2\varepsilon^{I}}{W'^{2}-u}, \nonumber \\
  E_{33}^{I} & = & \frac{1}{W'-W}+\frac{\varepsilon^{I}}{W'^{2}-u}\left(W-W'
               -2m_{N}-\frac{t-m_{\pi}^2}{W-m_{N}}\right), \nonumber \\
  E_{34}^{I} & = & \frac{1}{W'+W}+\frac{\varepsilon^{I}}{W'^{2}-u}\left(-W'-W
               +2m_{N}-\frac{t-m_{\pi}^2}{W+m_{N}}\right), \nonumber \\
  E_{41}^{I} & = & -\frac{2\varepsilon^{I}}{W'^{2}-u}, \nonumber \\
  E_{42}^{I} & = & -\frac{2\varepsilon^{I}}{W'^{2}-u}, \nonumber \\
  E_{43}^{I} & = & \frac{1}{W'+W}+\frac{\varepsilon^{I}}{W'^{2}-u}\left(-W'-W
               -2m_{N}-\frac{t-m_{\pi}^2}{W'-m_{N}}\right), \nonumber \\
  E_{44}^{I} & = & \frac{1}{W'-W}+\frac{\varepsilon^{I}}{W'^{2}-u}\left(W-W'
               +2m_{N}-\frac{t-m_{\pi}^2}{W'+m_{N}}\right). \nonumber
\end{eqnarray}

\section{Response functions}
The response functions discussed in this contribution may be expressed 
by the CGLN amplitudes as follows:
\label{resp_func}
\begin{eqnarray*}
  R_{T} & = & \mid{\cal F}_{1}\mid^{2}+\mid{\cal F}_{2}\mid^{2}
              +{\frac{1}{2}}\sin^{2}\theta\hspace{1mm}(
               \mid{\cal F}_{3}\mid^{2}
              +\mid{\cal F}_{4}\mid^{2}) \\ & & -\mbox{Re}\{
              2\cos\theta\hspace{1mm}{\cal F}_{1}^{\ast}{\cal F}_{2}
             -\sin^{2}\theta\hspace{1mm}({\cal F}_{1}^{\ast}{\cal F}_{4}
             +{\cal F}_{2}^{\ast}{\cal F}_{3}
             +\cos\theta\hspace{1mm}
              {\cal F}_{3}^{\ast}{\cal F}_{4})\}, \\[3mm]
  R_{T}^{n_{i}} & = & \sin\theta\hspace{1mm}\mbox{Im}\{
              {\cal F}_{1}^{\ast}{\cal F}_{3}
             -{\cal F}_{2}^{\ast}{\cal F}_{4} + \cos\theta\hspace{1mm}(
              {\cal F}_{1}^{\ast}{\cal F}_{4}
             -{\cal F}_{2}^{\ast}{\cal F}_{3})
             -\sin^{2}\theta\hspace{1mm}
              {\cal F}_{3}^{\ast}{\cal F}_{4}\}, \\[3mm]
  R_{T}^{n_{f}} & = & -\sin\theta\hspace{1mm}\mbox{Im}\{
               2{\cal F}_{1}^{\ast}{\cal F}_{2}
             +{\cal F}_{1}^{\ast}{\cal F}_{3}
             -{\cal F}_{2}^{\ast}{\cal F}_{4} + \cos\theta\hspace{1mm}(
              {\cal F}_{1}^{\ast}{\cal F}_{4}
             -{\cal F}_{2}^{\ast}{\cal F}_{3})
             -\sin^{2}\theta\hspace{1mm}
             {\cal F}_{3}^{\ast}{\cal F}_{4}\}, \\[3mm]
  R_{TT} & = & \sin^{2}\theta\hspace{1mm}({\frac{1}{2}}(
                  \mid{\cal F}_{3}\mid^{2}
                 +\mid{\cal F}_{4}\mid^{2}) + \mbox{Re}\{
                  {\cal F}_{1}^{\ast}{\cal F}_{4}
                 +{\cal F}_{2}^{\ast}{\cal F}_{3} 
                 + \cos\theta\hspace{1mm}
                  {\cal F}_{3}^{\ast}{\cal F}_{4}\}). \\[3mm]
\end{eqnarray*}

\acknowledgments

We would like to thank Prof.~G.~H\"ohler for very fruitful discussions
and the members of the A2 collaboration at Mainz for providing us with
their preliminary data, in particular J.~Arends, R.~Beck, F.~H\"arter, 
H.-P.~Krahn and H.~Str\"oher as well as G.~Anton from ELSA at Bonn.  This 
work was supported by the Deutsche Forschungsgemeinschaft (SFB~201).

\begin{table}[htbp]
    \caption{Compilation of the data included in our fit. The total number 
             of data points is 1751.}
  \begin{center}
    \leavevmode
    \begin{tabular}{lcrcr}
      reaction & observable & $E_{\gamma}$ [MeV] & number of data points
      & reference \\
      \hline
      $\gamma p\to\pi^{+}n$ & $d\sigma / d\Omega$ & 162--220 & 152 
      & \cite{Men77} \\
      & $d\sigma / d\Omega$ & 270--420 & 176
      & \cite{Kra96} \\
      & $d\sigma / d\Omega$ & 247--452 & 61 & \cite{Bue94} \\
      & $\Sigma$ & 270--420 & 176 & \cite{Kra96} \\
      & $\Sigma$ & 211--267 & 36 & \cite{Men77} \\
      & $T$ & 220--425 & 107 & \cite{Dut95} \\
      $\gamma p\to\pi^{0}p$ & $d\sigma / d\Omega$ & 160--180 & 
      224 & \cite{Fuc96} \\
      & $d\sigma / d\Omega$ & 270--420 & 107
      & \cite{Kra96} \\
      & $d\sigma / d\Omega$ & 210--425 & 82
      & \cite{Hae96} \\
      & $\Sigma$ & 270--420 & 107 & \cite{Kra96} \\
      & $T$ & 272--398 & 28 & \cite{Dut95} \\
      $\gamma n\to\pi^{-}p$ & $d\sigma / d\Omega$ & 190--420 & 189 
      & \cite{Car73} \\
      & $d\sigma / d\Omega$ & 190--230 & 20 & \cite{Ben58} \\
      & $d\sigma / d\Omega$ & 230--420 & 78 & \cite{Ben73} \\
      & $d\sigma / d\Omega$ & 250--420 & 158 & \cite{Fuj76} \\
      & $d\sigma / d\Omega$ & 194--270 & 50 & \cite{Bag88} \\
    \end{tabular}
    \label{tab:data}
  \end{center}
\end{table}
\begin{table}[htbp]
    \caption{Vector meson coupling constants from our fit compared  
             to the values used for the Bonn potential 
             \protect\cite{Mac87} and a dispersion theoretical 
             analysis of nucleon form factors \protect\cite{Mer95}.}
  \begin{center}
    \begin{tabular}{lrrrr}
    & $g_{\rho}^{{\cal V}}$ & $g_{\rho}^{{\cal T}}$ &
      $g_{\omega}^{{\cal V}}$ & $g_{\omega}^{{\cal T}}$ \\
    \hline
    this work & 4.85 & 15.69 & 6.78 & -1.67 \\
    Ref.~\cite{Mac87} & 3.24 & 19.81 & 15.85 & 0 \\
    Ref.~\cite{Mer95} & 1.99 & 12.42 & 20.86 & -3.41 \\
    \end{tabular}
    \label{tab:1}
  \end{center}
\end{table}
\begin{table}[htbp]
  \caption{The $S$ wave amplitude $E_{0+}$ at threshold 
           in units of $10^{-3}/m_{\pi}$.}
\vspace{0.4cm}
\begin{center}
\begin{tabular}{ccccc}
         & $\gamma p \rightarrow \pi^+n$ & $\gamma n\rightarrow \pi^-p$ 
     & $\gamma p \rightarrow \pi^{0}p$ & $\gamma n \rightarrow \pi^{0}n$ 
  \\ \hline
``LET''\cite{deB70}  & 27.5 & -32.0  & -2.4 & 0.4 \\
ChPT \cite{Ber96a,Ber96b} & $28.2 \pm .6$ & $-32.7 \pm .6$  & -1.16 & 2.13 \\
this work & 28.4$\pm$.2 & -31.9$\pm$.2  & -1.22$\pm$.16 & 1.19$\pm$.16 \\
experiment & $28.3 \pm .2$ \cite{Ada76}     
 & $-31.8 \pm .2$ \cite{Ada76} & -1.31$\pm$.08 \cite{Fuc96,Ber96}  & \\
\end{tabular}
\end{center}
\label{tab:2}
\end{table}

\begin{table}[htbp]
  \caption{Result of the speed plot}
\vspace{0.4cm}
\begin{center}
\begin{tabular}{lrrrr}
& $M_R$ (MeV)    & $\Gamma_R$ (MeV)& $r(10^{-3}/m_{\pi})$ & $\Phi$ (deg)
  \\
\hline
$M_{1+}$ & $1212 \pm 1$ & $ 99 \pm 2$     &  21.16  & -27.5       \\
$E_{1+}$ & $1211 \pm 1$ & $102 \pm 2$     &   1.23  & -154.7      \\
\end{tabular}
\end{center}
\label{tab:res}
\end{table}

\begin{table}[htbp]
  \caption{The ratio $R_{EM}$ and the helicity amplitudes $A_{\frac{1}{2}}$ 
           and $A_{\frac{3}{2}}$ for the $\gamma N\to\Delta$ transition. In 
           determining $A_{\frac{1}{2}}$ and $A_{\frac{3}{2}}$, for the 
           $\Delta(1232)$ a total width of 113 MeV has been assumed.}
\vspace{0.4cm}
\begin{center}
\begin{tabular}{lrrr}
& $R_{EM}$ (\%)    & $A_{\frac{1}{2}}$ ($10^{-3}/\sqrt{{\mbox GeV}}$)
&  $A_{\frac{3}{2}}$ ($10^{-3}/\sqrt{{\mbox GeV}}$) \\
\hline
global fit & -2.54$\pm$0.10 &  -131.2     &  -252.2      \\
local fit & -2.33$\pm$0.17 &  -129.4$\pm$1.3  &  -246.6$\pm$1.3      \\
\end{tabular}
\end{center}
\label{tab:hel}
\end{table}


\end{document}